\documentclass[reprint,superscriptaddress]{revtex4-1}
\usepackage{amsthm,amssymb}
\usepackage{mathrsfs}
\usepackage{graphicx}
\usepackage{caption3}
\usepackage{enumerate}
\usepackage{mathtools}
\usepackage{times, braket, color, bm}
\usepackage{multirow}
\usepackage{array} 
\usepackage{hyperref}
\usepackage{cleveref}
\hypersetup{colorlinks,linkcolor=myurlcolor,citecolor=myurlcolor,urlcolor=myurlcolor}

\definecolor{myurlcolor}{rgb}{0,0,0.7}
\newcommand{\proj}[1]{| #1\rangle\!\langle #1 |}

\def\be{\begin{equation}}
\def\ee{\end{equation}}
\def\bea{\begin{eqnarray*}}
\def\eea{\end{eqnarray*}}
\def\ot{\otimes}

\theoremstyle{plain}
\newtheorem{thrm}{\protect\theoremname}

\newtheorem{lem}[thrm]{Lemma}

\newtheorem{fact}[thrm]{Fact}
\newtheorem{coro}[thrm]{Corollary}
\newtheorem{rema}[thrm]{Remark}
\providecommand{\theoremname}{Theorem}
\newcommand{\inner}[2]{\langle #1 , #2\rangle}
\newcommand{\iinner}[2]{\langle #1 | #2\rangle}
\newcommand{\out}[2]{| #1\rangle\langle #2 |}
\DeclareMathOperator{\trace}{tr}
\newcommand{\ptr}[2]{\trace_{#1}({#2})}
\newcommand{\tr}[1]{\ptr{}{#1}}
\newcommand{\id}{\mathbb{I}}

\newcommand*{\myproofname}{Proof}

\newcommand{\conv}{\operatorname{conv}}

\newcommand{\diag}{\operatorname{diag}}
\newcommand{\lamax}{\lambda_{\max}}
\newcommand{\half}{\tfrac12}
\newcommand{\Pos}{\mathsf{Pos}}
\newcommand{\MU}{\mathsf{MU}}
\makeatother

\def\cC{\mathcal{C}}\def\cD{\mathcal{D}}\def\cE{\mathcal{E}}
\def\cH{\mathcal{H}}
\def\cM{\mathcal{M}}\def\cN{\mathcal{N}}
\def\cR{\mathcal{R}}\def\cS{\mathcal{S}}\def\cT{\mathcal{T}}

\def\rD{\mathrm{D}}
\def\rF{\mathrm{F}}

\def\D{\textsf{D}}

\theoremstyle{definition}

\theoremstyle{remark}

\graphicspath{{figs/}}

\begin{document}

 \author{Sunho Kim}
 \email{kimsunho81@hrbeu.edu.cn}
 \affiliation{School of Mathematical Sciences, Harbin Engineering University, Harbin 150001, China}

  \author{Chunhe Xiong}
 \email{Corresponding author: xiongchunhe@csu.edu.cn}
\affiliation{School of Mathematics and Statistics, HNP-LAMA, Central South University, Changsha 410083, China}

 \author{Junde Wu}
 \email{Corresponding author: wjd@zju.edu.cn}
 \affiliation{School of Mathematical Sciences, Zhejiang University, Hangzhou 310027, China}

\title{Quantum Defect Analysis and Target-Orbit-Directed Correction via the Purpose-Oriented Framework: Applications to Entanglement Repairability and Teleportation Activation}
\begin{abstract}
The fundamental task of quantum resource theory is to quantify the advantages that quantum resources provide in specific tasks and to establish their operational meanings. However, conventional theories suffer from structural ambiguities in the set of free states, rendering quantification complex and inherently restricting their applicability in target-oriented missions where the free set is ill-defined or certain resource properties remain inactive. To improve upon these limitations and systematize the approach, this work introduces a purpose-oriented framework that shifts the reference point from ``free'' to ``maximally efficient.'' Within this framework, we employ a deficiency measure that quantifies the system's loss and deficiency, enabling the classification of mixed resource states and the experimental estimation of noise constants. We further show that this measure serves as a key indicator for determining quantum-error-correction thresholds and predicting the algorithmic performance of Hadamard gates.

Furthermore, we define an operational quantity called ``repair capacity'' to demonstrate how the deficiency measure functions as a metric for mission success in environments where the free set is ambiguous. We prove that this capacity equals a one-sided spectral separation of the pulled-back effects from the target hull, thereby exposing the mixed-unitary channel boundary. Building on this theoretical rigor, we derive the affine Bloch translation law for qubit channels and the odd-even separation for Werner--Holevo channels.

In particular, for one-sided qubit generalized amplitude damping (GAD) channels, optimizing over all inputs below the classical teleportation limit reveals the finite-temperature onset $z_\beta=1/\sqrt{2}$ and yields the optimal interaction time. This demonstrates that establishing the purpose-oriented framework is not merely a mathematical alternative, but rather enables the systematic integration of non-random-unitary dynamics, algorithmic calibration, and dissipative task activation.
\end{abstract}
\maketitle

\section{Introduction}
\label{sec:intro}

Quantum resource theories (QRTs) have provided a versatile framework for studying various phenomena in quantum information, including entanglement~\cite{Bennett, Horodecki1, Hou, Roa}, coherence~\cite{Plenio, Luo}, nonlocality~\cite{Schaetz}, and thermodynamics~\cite{Gour}.
Foundational treatments of general resource quantification further unify these perspectives~\cite{Regula}.
Conventional state-based approaches define free states as a convex and closed set, quantifying operational advantages through resource measures satisfying faithfulness, monotonicity, and convexity~\cite{Chitambar}.
Various quantification methods have been developed, such as robustness~\cite{Vidal, Harrow, Napoli}, distance-based measures~\cite{Vedral, Baumgratz}, and distillation~\cite{Bennett, Winter, Bravyi}.
Specific measures characterize physical tasks offering explicit advantages over free states, thereby endowing a given resource with operational meaning~\cite{Zheng, Takagi, Ducuara, Uola}.

However, when rigorously characterizing the operational advantages for specific quantum tasks, the conventional approach of defining ``free states'' based solely on the ease of preparation exhibits severe ambiguity and limitations.
Traditional quantum resource theories (QRTs) assume that free states form a convex and closed set, upon which resource measures are constructed.
Nevertheless, in practical physical scenarios, the set of states that provide no quantum advantage often exhibits a severely nonconvex structure, leaving regions that cannot be explained by conventional convex-optimization-based approaches.
For instance, bound entangled states do not enhance the fidelity of standard teleportation protocols~\cite{Horodecki5, Linden}.
Conversely, states that are useless for Grover search may still possess nonzero coherence~\cite{Biham}.
These phenomena clearly demonstrate the disparity between the mathematical properties of states and their actual operational usefulness.
To overcome these fundamental limitations, recent studies have begun developing target-oriented resource theories that completely dispense with convexity requirements on free states~\cite{Kuroiwa, Kuroiwa2, Salazar}.
This represents a significant paradigm shift, moving the research focus from state preparation to actual task performance.

Meanwhile, similar issues are observed in the field of evaluating the operational performance of quantum channels.
Previous studies have predominantly relied on channel symmetries, particularly specific mathematical properties such as unitality, to analyze performance boundaries.
In entanglement-assisted teleportation, for example, it is well known that arbitrary unitary dynamics cannot increase the fully entangled fraction (FEF) of a shared state~\cite{BadziagHorodeckiHorodeckiHorodecki2000}.
Subsequent studies revealed that it is not unitality per se, but rather mixed-unitarity that determines the critical performance boundary, suggesting deep mathematical connections to operator distance measures such as the Choi and diamond distances~\cite{MendlWolf2009, GirardLeungLevickLiPaulsenPoonWatrous2022, YuDuanXu2017}.
Furthermore, research on the characterization of high-dimensional Werner--Holevo channels~\cite{Karimipour2024} and local dissipative effects in thermal environments~\cite{Bandyopadhyay2002, Yeo2008, LiuLi2025} has been actively conducted.
Nevertheless, these studies remain confined to specific symmetry conditions or particular environmental models, limiting their applicability to general-purpose quantum devices.
Considering that the utility of an operation is only determined after the operational goal of the device is specified~\cite{Kuroiwa}, a unified framework that rigorously quantifies the operational efficiency of quantum channels---without relying on the ambiguous concept of ``free'' states---remains absent.

In this work, we propose such a framework.
In Sec.~\ref{sec:2}, we propose a purpose-oriented deficiency quantification protocol that eliminates the concept of ``freeness'' entirely.
The core of this protocol is to circumvent the ambiguity of free states by defining the ``distance'' between the given experimental target (e.g., maximally resourceful states or states of maximal operational efficiency) and an arbitrary input state as a deficiency loss function.
Through this framework, we derive the following results.
In Sec.~\ref{sec:3}, we introduce a geometric deficiency metric and show that it rigorously characterizes operational weakness in sub-channel quantum discrimination tasks (Sec.~\ref{sec:3-a}).
Furthermore, we demonstrate that this deficiency measure is highly suitable for quantifying noise in quantum channels (Sec.~\ref{sec:3-b}).
In Sec.~\ref{sec:4}, to show that the deficiency measure functions as an indicator of task success, we define an operational quantity termed repair capacity.
This is defined as the one-sided spectral separation of the effect pulled from the target hull.
From this definition, we derive two subsidiary results:
First, analysis of the repair capacity proves that a channel being mixed-unitary is equivalent to the impossibility of reducing the system's entanglement deficiency via local operations.
This provides a clear boundary regarding the repairability of multi-dimensional quantum entanglement. We apply this framework to the affine Bloch transformation laws of qubit channels and the odd-even separation of Werner--Holevo channels (Sec.~\ref{sec:4-a}).
Second, by performing a global optimization for a one-sided qubit Generalized Amplitude Damping (GAD) channel, we derive a lower critical parameter than the conventional local threshold and analytically present a temperature-dependent onset value. (Sec.~\ref{sec:4-b}).

These results demonstrate that establishing the purpose-oriented framework is not merely a mathematical alternative to convex resource monotones,
but an effective and operational approach that bridges non-random-unitary dynamics to algorithmic calibration and dissipative
task activation.
Our work paves the way for a more flexible and task-oriented characterization of quantum channels, with direct implications for quantum communication, error mitigation, and the design of noise-resilient quantum protocols.

\section{Quantum resource theory of deficiency for the maximal resource states}\label{sec:2}

Building on the observation that quantum advantage is commonly gauged by fidelity to maximal resources, we propose a novel approach.
Teleportation uses Bell-state fidelity \cite{Bennett2,Horodecki3,Horodecki4,Bouwmeester}, while Grover's algorithm relies on the coherence fraction \cite{Zhou}. Because prepared states rarely attain maximality, formulating a target-oriented quantitative framework relative to optimal resources
is practically significant. Such scenarios---lacking a preexisting resource notion---are ubiquitous in target-oriented missions. Our
purpose-oriented framework nonetheless yields a rigorous basis for channel characterization and resource analysis sans a free-state
manifold.

As a concrete illustration, consider $\mathcal{D}(\mathcal{H})$ denote the density operators on a finite-dimensional Hilbert space $\mathcal{H}$. A target specification is a nonempty compact family $\mathcal{T}\subset\{E:0\leq E\leq I\}$ of acceptance effects. Each $E$ is the pass outcome of a two-outcome test $\{E,I-E\}$; this family may contain a single desired state, a group orbit, a code projector, or a finite menu of acceptable outputs. Defining its compact convex hull $\mathcal{C}=\conv\mathcal{T}$ and the target affinity
\begin{equation}
 f_{\mathcal{T}}(\rho)=\max_{E\in\mathcal{T}}\tr{E\rho}
 =\max_{\omega\in\mathcal{C}}\tr{\omega\rho},
\end{equation}
we obtain the optimal pass probability allowed by the target menu. We then define the states achieving $f_{\mathcal{T}}(\rho)=1$ as the
maximally efficient (or target) states for this task, formally denoted by
\begin{equation}
 \mathcal{S}_{\mathcal{T}}^{\max} \coloneqq \{\,\rho\in\mathcal{D}(\mathcal{H})\mid f_{\mathcal{T}}(\rho)=1\,\}.
\end{equation}
In such scenarios the free manifold is either undefined or loses its operational significance.

To treat these scenarios systematically, this work first focuses on mathematically establishing the purpose-oriented framework itself. We
invert standard axioms to present a deficiency-based framework. This approach not only applies to nonconvex or unspecified free sets but
also aligns better with experimental realities where access to ideal maximal resources is rare. By reframing the problem in terms of
relative deficiency, we provide a flexible and operational method for evaluating resource quality across diverse quantum information processing
scenarios.

Based on this perspective, we propose a new approach for measuring the degree of deficiency through conditions that contrast with the fundamental
properties of quantum resources in conventional quantum resource theory. Crucially, unlike standard resource theories, our framework starts from
a setting where free states and free operations are not considered. In this basic setting, the deficiency for maximally efficient (resource)
states is defined by a function $\rD$ that satisfies the following two basic conditions:

(D1) (Faithful): $\rD(\sigma)\geq 0$, and $\rD(\sigma)=0$ if and only if $\sigma \in \cS^{\max}$, where $\cS^{\max}$ is the set of all maximally efficient (resource) states $\sigma$;

(D2) (Concave): $\rD(\sum_iq_i\sigma_i) \geq \sum_iq_i\rD(\sigma_i)$.

\medskip
We define a function $\rD$ satisfying these two basic conditions as a deficiency measure.

These two conditions indicate that as the measured indicator increases, the deficiency relative to the maximally efficient states required for
target task accomplishment grows (D1), and further guarantee that the resource efficiency in quantum tasks is consistent with the concave
decreasing trend of the deficiency (D2). This reveals that the concept of resource convexity extends beyond a mere mathematical property to
an intuitive correspondence between the degradation of task performance and the increase of deficiency.
Because we dispense with free operations, the complications regarding monotonicity for mixed states---such as those arising in conventional measures due to relative phase sensitivity---are naturally avoided.
The absence of a monotonicity constraint reflects the fact that, in task-oriented settings, the operational advantage is quantified solely by the distance from the target family, rather than by transformations under a fixed set of free channels.

In situations where free operations are specifically taken into account, a nondecreasing monotonicity requirement for the deficiency may be imposed.
For this purpose, we introduce the following additional monotonicity conditions:

(D3a) (Nondecreasing) monotonicity for pure states under any free operation $\Phi$: $\rD(\Phi(\out{\psi}{\psi})) \geq \rD(\out{\psi}{\psi})$,\\
or (D3b) monotonicity under selective measurement $\{K_n\}$ : $\sum_np_n\rD(\out{\psi_n}{\psi_n})\geq \rD(\out{\psi}{\psi})$, where $\ket{\psi_n} = K_n\ket{\psi}/\sqrt{p_n}$ with $p_n = \tr{K_n\out{\psi}{\psi} K_n^\dagger}$;

(U-D3a) (Nondecreasing) universal monotonicity for all states under any free operation $\Phi$: $\rD(\Phi(\rho)) \geq \rD(\rho)$,\\
or (U-D3b) monotonicity under selective measurement $\{K_n\}$ : $\sum_np_n\rD(\rho_n)\geq \rD(\rho)$, where $\rho_n = K_n\rho K_n^\dagger/p_n$ with $p_n = \tr{K_n\rho K_n^\dagger}$.

\medskip

If a deficiency measure further satisfies the monotonicity conditions (D3), we call it a monotonic deficiency.
Furthermore, if it satisfies the universal monotonicity conditions (U-D3), we define it as a universal monotonic deficiency.

To illustrate with coherence as an example: quantum coherence is typically quantified by specific functions of the absolute values of off-diagonal elements in the density matrix with respect to the reference basis.
These measures include the robustness measure \cite{Napoli}, $l_1$-norm measure \cite{Baumgratz}, relative entropy measure \cite{Winter}, and geometric measure \cite{Streltsov2} of coherence.
This means that states with identical absolute values of off-diagonal elements possess equal coherence.
However, the pure-state nature of maximal resource states implies that even when two quantum states have equal absolute values of off-diagonal elements, their resource deficiency may differ due to relative phase differences.
This phase sensitivity suggests that monotonicity might not hold for certain mixed states, but since we are dealing with the basic setting where free operations are not considered, this lack of monotonicity is not a limitation of our framework but rather a natural consequence.

For pure states, however, the phase dependence cancels out precisely because maximal resource states are also pure--this implies that the deficiency is unaffected by phase differences. This fundamental property ensures non-decreasing monotonicity for pure states, which aligns with conventional quantum resource theory.

\section{Geometric approach to resource deficiency quantification}\label{sec:3}

The deficiency quantification approach proposed in this study takes the maximal resource set as its reference. Furthermore, the fact that the maximal resource set consists exclusively of pure states implies that certain measures--including robustness-based measures--fail to satisfy the requirements of this framework for quantifying relative deficiency.
This occurs because pure states cannot be obtained through any convex combination of quantum states. We therefore propose the use of a geometric measure to quantify resource deficiency in this context.

Given a state $\rho$, we define the geometric measure of deficiency relative to maximal resource states as
\be
\rD_g(\rho) = \min_{\sigma\in\cS^{\max}}\Big\{1-\rF(\sigma,\rho)\Big\}
\ee
where the fidelity $\rF(\sigma,\rho) = \|\sqrt{\sigma}\sqrt{\rho}\|_1^2$ for two positive semidefinite operators $\sigma, \rho$.

Furthermore, since all maximal resource states are pure, the expression simplifies to $\rD_g(\rho) = \min_{\sigma\in\cS^{\max}}\{1-\inner{\Pi_\sigma}{\rho}\}$, where $\Pi_\sigma$ denotes the projection operator onto the eigenstates with nonzero eigenvalues of $\sigma$.
We next check that this geometric function $\rD_g$ is suitable for measuring resource deficiency for coherence and entanglement, respectively
(see the proofs of the following two theorems in Appendix~\ref{Appendix A}).

\begin{thrm}\label{thm1}
We define
\be\label{eq2}
\rD^C_g(\rho) = \min_{\sigma\in\cC^{\max}}\Big\{1-\rF(\sigma,\rho)\Big\}
\ee
where $\cC^{\max}$ is the set of all maximal coherent states in $\cD(\cH)$.
Then, $\rD^C_g$ is a monotonic measure of coherence deficiency.
\end{thrm}

\begin{thrm}\label{thm2}
We define
\be\label{eq3}
\rD^E_g(\rho) = \min_{\sigma\in\cE^{\max}}\Big\{1-\rF(\sigma,\rho)\Big\}
\ee
where $\cE^{\max}$ is the set of all maximal entangled states in $\cD(\cH_A\ot \cH_B)$.
Then, $\rD^E_g$ is a monotonic measure of entanglement deficiency.
\end{thrm}

Here, for a quantum state $\rho$, if there exists a appropriate maximal resource state $\ket{\psi}_{max}=\frac{1}{\sqrt{d}}\sum_ie^{i\theta_i}\ket{\psi_i}$ such that the phase structure of $\rho$ is completely canceled by $\ket{\psi}_{max}$, that is, $\rD_g(\rho)= 1-\frac{\sum_{i,j}|\rho_{ij}|}{d}$ where $\rho_{ij} = \iinner{\psi_j|\rho}{\psi_i}$, then we define the phase structure of $\rho$ as consistent. Therefore, every pure state has a consistent phase structure.

Verifying whether the geometric deficiency measure $\rD_g$ exhibits universal monotonicity (U-D2b) for arbitrary quantum resources is nontrivial. However, when $\dim(\cH) = 2$ for coherence deficiency $\rD^C_g$ these quantities can be rigorously demonstrated to satisfy universal monotonicity as valid deficiency measures (proofs of these results are provided in Appendix \ref{Appendix B}).

\begin{thrm}\label{thm3}
For $\dim(\cH) = 2$, $\rD^C_g$ becomes a measure of coherence deficiency satisfying universal monotonicity (U-D2b).
\end{thrm}

\begin{rema} \label{Violation of universal monotonicity in entanglement}
In general, universal monotonicity does not hold for entangled mixed states.
Any mixed state can undergo a reduction in mixedness under certain local operations.
For instance, consider the local operation $R_A \otimes R_B$ defined by Kraus operators $K_0=|0\rangle\langle 0|$ and $K_1=|0\rangle\langle 1|$.
When applied to the maximally mixed state $\rho = I/4$, this operation transforms it into the pure separable state $|00\rangle\langle 00|$.
Consequently, the maximal fidelity increases from $1/8$ to $1/2$, implying that $\rD^E_g$ decreases from $7/8$ to $1/2$.
This counterexample demonstrates that for entangled mixed states, the inequality $\sum p_{n,m}\rD^E_g(\rho_{n,m}) \ge \rD^E_g(\rho)$ may be violated due to the potential reduction of mixedness via local projections.

Notably, this behavior distinguishes entanglement from two-dimensional coherence.
In the case of two-dimensional coherence, even if the purity of a free state increases under free operations (Incoherent Operations, IO)---for example, $I/2 \to |0\rangle\langle 0|$---the maximal fidelity does not increase.
This, combined with the existence of a maximally coherent state that preserves relative phase information, guarantees universal monotonicity.
In contrast, for entanglement, purity increase under certain separable operations can lead to an increase in maximal fidelity, and for this reason, universal monotonicity cannot be guaranteed.
\end{rema}

Geometric measures typically capture only a single characteristic of a resource (e.g., the largest Schmidt coefficient) and are therefore limited.
In contrast, the geometric measure of deficiency for maximal resources---which is defined via the standard fidelity---considers the full distribution of all Schmidt coefficients as well as phase-structure inconsistency in mixed-state scenarios, thereby providing more complete structural information. Beyond the known undistillability of PPT entangled states, we identify that degradation of resource efficiency can also originate from the decline in relative similarity to maximal entanglement caused by phase-structure inconsistency.

In particular, phase-structure inconsistency cannot be detected through comparison with the set of free states; however, in two-dimensional pure states (where the universal monotonicity of coherence deficiency is rigorously established), it cancels out, preserving the monotonicity and consistency of standard resource theories.
We have confirmed that for entanglement, the universal monotonicity does not generally hold even in low dimensions.
Consequently, rather than proposing a new quantifier, our work offers an additional interpretation of the fidelity-based deficiency.
This suggests that, although this geometric approach does not completely delineate the boundary of efficiency, it can serve as a refined analytical tool for reinterpreting and classifying mixed states whose resource properties remain latent under certain operations---such as PPT entangled states---compared to existing methods.

We demonstrate how the geometric measure of deficiency can be employed as an operational handle to quantify operational disadvantages and noise in quantum tasks.

\subsection{Operational disadvantage for maximal resources in subchannel discrimination}\label{sec:3-a}
Our ultimate goal is to establish an indicator for the operational disadvantages of quantum states in subchannel discrimination and to relate this indicator to resource deficiency. As a first step, we refine the advantage indicator proposed in \cite{Takagi} to more reasonably capture operational disadvantages.

In previous work, the relative advantage of quantum states in subchannel discrimination was quantified using the maximum ratio of success probabilities across all possible strategies. However, this approach has a limitation: while it maximizes the success probability ratio, it does not guarantee sufficiently high absolute success probabilities for the quantum states involved.

To address this issue, we introduce an improved indicator for operational disadvantages. Rather than considering all possible strategies, we compute the relative ratio only over strategies that achieve the maximum success probability for maximal resource states. This ensures a more meaningful measure of disadvantage while maintaining high success probabilities. For any given strategy $(\{\Psi_i\}, \{M_i\})$, when the success probability for subchannel discrimination is given by $P_{succ}(\{\Psi_i\}, \{M_i\}, \sigma) = \sum_i\tr{M_i\Psi_i(\sigma)}$, the indicator is expressed as follows:
\be\label{def4}
\max_{\sigma\in \cS^{\max}}\min_{\Omega_\sigma}\frac{P_{succ}(\{\Psi_i\}, \{M_i\}, \rho)}{P_{succ}(\{\Psi_i\}, \{M_i\},\sigma)}
\ee
where $\Omega_\sigma = \big\{(\{\Psi_i\}, \{M_i\})|P_{succ}(\{\Psi_i\}, \{M_i\}, \sigma) = 1\big\}.$
Under strategies that maximize the success probability for maximal resource states (as defined in Eq. (\ref{def4})), the minimum ratio between the success probabilities obtained using two quantum states quantifies the relative disadvantage of a given state $\rho$ compared to a maximal resource state $\sigma$. By maximizing these ratios over all possible maximal resource states, we determine the overall operational disadvantage of $\rho$ relative to the entire set of maximal resource states.

This indicator exhibits an inverse relationship with operational disadvantage: higher values correspond to smaller disadvantages, while lower values indicate greater disadvantages. Importantly, through strategies that yield this disadvantage value, we still guarantee the maximum success probability for the maximal resource state that produces the maximum ratio.

In our proposed indicator, we exclusively consider strategies that maximize the success probability for maximal resource states. Since maximal resource states are typically pure states, we establish conditions under which strategies guarantee a maximum success probability of $1$. As demonstrated in Appendix \ref{Appendix C}, infinitely many strategies satisfy these conditions.

Moreover, our results reveal a significant connection: the operational disadvantages arising from quantum states in the subchannel discrimination framework can be precisely characterized by a geometric measure of resource deficiency. (See the proof in Appendix \ref{Appendix D})

\begin{thrm}\label{thm7}
For any $\rho\in \cD(\cH)$,
\be\label{5}
\max_{\sigma\in \cS^{\max}}\min_{\Omega_\sigma}\frac{P_{succ}(\{\Psi_i\}, \{M_i\}, \rho)}{P_{succ}(\{\Psi_i\}, \{M_i\},\sigma)} = 1-\rD_g(\rho).
\ee
\end{thrm}

These results demonstrate that for maximal resource sets consisting exclusively of pure states, the operational disadvantages of arbitrary quantum states in subchannel discrimination can be quantified using a geometric measure of resource deficiency. Unlike robustness measures, which are typically employed to quantify operational advantages in closed convex resource theories, our geometric approach successfully quantifies operational disadvantages even in cases where robustness measures cannot be defined for resource deficiencies. Furthermore, existing research has been limited to characterizing resource advantages only when the boundaries of closed convex sets are clearly defined. This poses particular constraints in applying conventional methodologies to PPT states, which have no operational efficiency and lack clearly defined boundaries. However, since most quantum resources exhibit maximal resource states with distinct boundaries, employing deficiency measures can effectively improve upon these limitations.

It should be noted with caution that these results are applicable to any quantum resource theory where all maximal resource states are pure states. Consequently, due to their dependence on the specific structure of maximal resource states, these results cannot be universally extended to arbitrary closed sets.

\subsection{Estimating quantum circuit noise via deficiency measures}\label{sec:3-b}
In quantum information processing, the core resource of parallelism is based on the creation of superposition states. These superposition states are typically determined by the accuracy of prepared initial states and superposition-generating gates (e.g., Hadamard gates), and the noise environment of the gates directly impacts the quality of the superposition. Therefore, quantifying the noise characteristics of quantum gates is a central challenge in quantum computing implementation. Gate noise can be decomposed into two components: a fixed instability constant $\epsilon_0$ (the fixed error component of the quantum circuit) and a time-dependent coupling constant $\gamma k$ (the error component proportional to the gate implementation time). From this, the total noise constant $\epsilon_G$ of a gate $G$ is expressed as $\epsilon_G = \epsilon_0 + \gamma kT_G$, where $T_G$ is the execution time for gate $G$. The purpose of this study is to estimate the noise constant $\epsilon_{\textmd{H}}$ of the Hadamard gate in an $n$-qubit system, and the noise characteristics can be characterized through the fidelity between the ideal maximal superposition state and the resulting state $\rho_{\textmd{noise}}$ in a noisy environment. The deficiency measure $\rD_g(\rho_{\textmd{noise}})$ is defined as $\rD_g(\rho_{\textmd{noise}}) = 1 - \textmd{F}_{\textmd{max}}$, where it is reasonable to regard $\textmd{F}_{\textmd{max}}$ as the fidelity between the state $\rho_{\textmd{noise}}$ and the original ideal target state.
Considering that the ideal $n$-qubit maximal superposition state is $\ket{+}^{\otimes n}$, we can set $\textmd{F}_{\textmd{max}}= \textmd{F}(\out{+}{+}^{\otimes n}, \rho_{\textmd{noise}})$ . The resulting state generated by the noise constant $\epsilon_{\textmd{H}}$ is expressed as the tensor product of noise channels acting independently on each qubit: $\rho_{\textmd{noise}} = \otimes_{i=1}^n \Phi_{\epsilon_{\textmd{H}}}(\out{+}{+}_i)$, where $\Phi_{\epsilon_{\textmd{H}}}(\out{+}{+}) = (1-\epsilon_{\textmd{H}})\out{+}{+} + \epsilon_{\textmd{H}} \id/2$. From this relationship, the noise constant estimation formula
\begin{equation}
2[1 - \sqrt[n]{1-\rD_g(\rho_{\textmd{noise}})}]= 2[1 - \sqrt[n]{\textmd{F}_{\textmd{max}}}] = \epsilon_{\textmd{H}}
\end{equation}
can be derived. Furthermore, for sufficiently small $\epsilon_{\text{H}}$, the relationship can be simplified to the following approximate form:
\begin{equation}
\frac{2\rD_g(\rho_{\textmd{noise}})}{n}=\frac{2(1 - \textmd{F}_{\textmd{max}})}{n} \approx \epsilon_{\textmd{H}}.
\end{equation}
This relation shows that the noise in a quantum system can be characterized through the deficiency measure $D_g(\rho_{\text{noise}})$ proposed in this study. Specifically, the single-qubit gate noise constant $\epsilon_{\textmd{H}}$ is approximately proportional to the normalized deficiency $2\rD_g/n$.

\begin{figure}[!t]
\centering
\includegraphics[height=.22\textheight]{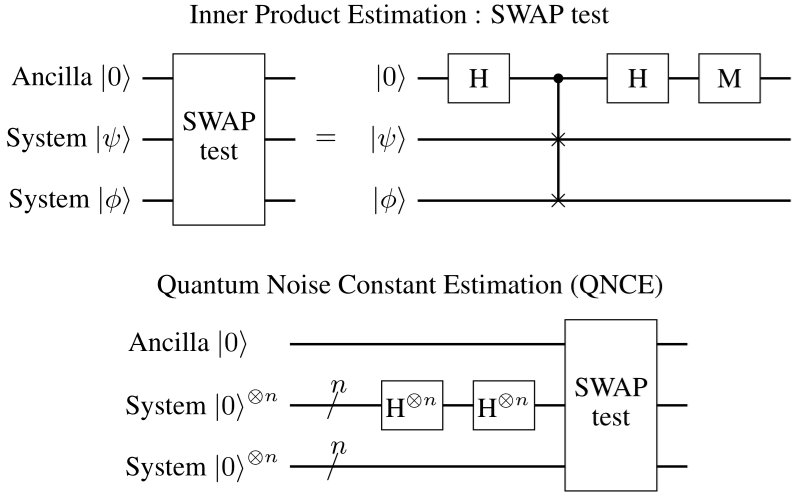}
\caption{\label{fig:2}  Quantum Noise Constant Estimation (QNCE) via quantum deficiency computed using SWAP test algorithms within a given circuit environment.}
\end{figure}

This work demonstrates that the noise constant characterizing a noisy environment can be estimated via deficiency measures. While the deficiency is defined as the fidelity between a noisy state and the ideal maximal superposition state, preparing such an ideal state under the same noise conditions is experimentally challenging. We show that the deficiency can instead be measured via the fidelity between two noisy states: one prepared from the initial state $\ket{0}^{\otimes n}$ and the other obtained by applying the Hadamard gate twice, thereby providing the desired deficiency measure. For this purpose, we propose the following practical procedure for determining the deficiency measure.

The inner product between the states resulting from these two noisy evolutions can be estimated via a quantum overlap estimation algorithm (e.g., the SWAP test \cite{Buhrman}, see Fig.~\ref{fig:2}). The estimated overlap directly yields the deficiency measure, which in turn furnishes an estimate of the overall noise constant $\epsilon_{\textmd{H}}$.

We introduce a concrete methodology for quantifying quantum gate noise characteristics. The methodology theoretically links deficiency measures from quantum resource theory with noise constants and presents an implementation suitable for experimental estimation. Specifically, we analyze the change in state fidelity under repeated application of a noisy channel and derive explicit relations for two distinct cases--with and without crosstalk in multi-qubit systems--that enable the measurement of deficiency relative to maximal superposition states. The proposed approach relies on inner-product measurements for experimental estimation, quantifies the statistical fluctuations of the measured data through error propagation, and provides confidence intervals for the estimated noise constant and deficiency measure. Based on this, we analyze the error margins of the estimated deficiency measure and the relative error of the noise constant for sample sizes $N_s = 10^3$ and $10^4$ and for qubit numbers $n = 1,2,\dots,8$ (the detailed scheme and results are provided in Appendix~\ref{Appendix E}).

This study proposes a methodology for experimentally estimating quantum-gate noise constants by linking them with deficiency measures. The approach not only enables the quantification of noise characteristics for Hadamard gates but can also be extended to analyze the noise properties of more general quantum gates. For example, the same framework can be applied to characterize the noise of elementary entangling circuits such as the Bell-state preparation circuit, which consists of a Hadamard gate followed by a CNOT gate. Noise constants serve as key indicators for determining quantum error-correction thresholds and predicting the performance of quantum algorithms. Therefore, the relationship established between noise constants and deficiency measures confirms that deficiency values estimated via the proposed methodology can likewise be employed as reliable performance-prediction indicators.

\section{Target repair and order duality}
\label{sec:4}

Following the purpose-oriented framework established above, the setting where the free-state manifold is not independently specified or treated
as a primary object is naturally extended. Even in the absence of a predefined free-state manifold, the target-repair still provides
a rigorous operational meaning for a channel's ability to reduce entanglement deficiency.

To recapitulate briefly: for density operators on $\mathcal{D}(\mathcal{H})$, a target specification is given by a
nonempty compact family $\cT\subset\{E:0\leq E\leq I\}$ of acceptance effects. The target affinity $f_{\mathcal{T}}(\rho)=\max_{E\in\mathcal{T}}\tr{E\rho}$
represents the optimal pass probability allowed by the target menu, and states satisfying $f_{\mathcal{T}}(\rho)=1$ are defined as the maximally
efficient states $\mathcal{S}_{\mathcal{T}}^{\max}$ for this task.
These are the states perfectly aligned with the acceptance menu $\cT$.

While the set $\cS_{\cT}^{\max}$ can be empty if the input is too noisy or the specification $\cT$ is too restrictive (e.g., a single pure state unreachable by the given dynamics), it is generally nonempty in our target-repair architecture. Specifically, when $\cT$ is defined as a group orbit or a code subspace, the repair operation is designed to pull the state into this orbit, guaranteeing the existence of maximally efficient states and ensuring $f_{\cT}(\rho)=1$ is attainable.

It is important to note that, by the very nature of this task, it is not appropriate to define a set of free states (as in standard quantum resource theories). In typical resource theories, free states form a fixed, background convex set (e.g., separable or incoherent states) that are considered "valueless" and serve as the baseline for resource quantification. However, our framework is inherently goal-oriented and asymmetric: the target menu $\cT$ acts as an attractor that the protocol attempts to reach, rather than a forbidden or valueless background. Since the operational objective is to maximize the affinity $f_{\cT}(\rho)$ via active repair, partitioning states into "free" and "resourceful" categories a priori does not align with the adaptive, target-pulling character of the architecture. Consequently, the deficiency $1-f_{\cT}(\rho)$ functions strictly as a task loss rather than a resource monotone.

To quantify how far a given state $\rho$ is from the maximally efficient set $\cS_{\cT}^{\max}$, we evaluate the deficiency with respect to the target specification, defined as
\begin{equation}
 \rD_{\cT}(\rho)=1-f_{\cT}(\rho).
\end{equation}
It is crucial to emphasize that $\rD_{\cT}$ functions strictly as a task loss. Unlike standard resource monotones in quantum resource theories (e.g., entanglement monotones or coherence measures), which are postulated in advance based on axiomatic properties such as convexity or asymptotic continuity, $\rD_{\cT}$ is defined a posteriori by the specific operational requirement of the target menu $\cT$. $\rD_{\cT}$ should not be regarded as a universal measure of quantumness or a task-independent resource monotone. Rather, it is a contextual, goal-oriented metric: it evaluates the immediate penalty (or probability of failure) incurred when $\rho$ is subjected to the acceptance test $\{E, I-E\}$. Consequently, minimizing $\rD_{\cT}$ is equivalent to maximizing the pass probability, driving the state-repair protocol toward the target orbit without assuming any pre-defined resource hierarchy.
The faithfulness with respect to the target state set $\cS_{\cT}^{\max}$ and the concavity of the deficiency measure $\rD_{\mathcal{T}}$ follow identically to the proof of the geometric deficiency. Here, we only verify the product properties of $\rD_{\mathcal{T}}$ for multiple systems.

When systems $A$ and $B$ are in a product state and the target also admits a product structure, we apply the property of $f$:
\begin{equation}
 f_{\mathcal{T}_A \otimes \mathcal{T}_B}(\rho_A \otimes \rho_B) = f_{\mathcal{T}_A}(\rho_A) f_{\mathcal{T}_B}(\rho_B).
\end{equation}
Substituting this into the definition of $\rD$ yields
\begin{equation}
 \rD_{\mathcal{T}_A \otimes \mathcal{T}_B}(\rho_A \otimes \rho_B) = 1 - f_{\mathcal{T}_A}(\rho_A) f_{\mathcal{T}_B}(\rho_B).
\end{equation}
Letting $f_{\mathcal{T}_A}(\rho_A) = 1 - \rD_A$ and $f_{\mathcal{T}_B}(\rho_B) = 1 - \rD_B$, we obtain
\begin{align}
 \rD_{A \otimes B} &= 1 - (1 - \rD_A)(1 - \rD_B) \nonumber \\
 &= 1 - (1 - \rD_A - \rD_B + \rD_A \rD_B) \nonumber \\
 &= \rD_A + \rD_B - \rD_A \rD_B.
\end{align}
This implies that the deficiency measure decomposes additively (specifically, sub-additively) across subsystems under independent measurements. Consequently, for large multi-copy systems, it mathematically guarantees that the optimization can be decoupled into independent optimizations over each small subsystem.

The dynamical question begins with a finite quantum instrument $\{\Phi_i\}_{i=1}^n$. Each $\Phi_i$ is completely positive and trace nonincreasing, and $\sum_i\Phi_i$ is trace preserving. For input $\rho$, write $p_i=\tr{\Phi_i(\rho)}$ and $\rho_i=\Phi_i(\rho)/p_i$ when $p_i>0$. We quantify the average repair capacity by evaluating the optimal deficiency gap as follows:
\begin{equation}
\Delta \rD_{\cT}(\{\Phi_i\}) \coloneqq
 \max_\rho\left[\rD_{\cT}(\rho) - \sum_i p_i\rD_{\cT}(\rho_i)\right].
\end{equation}
It is nonnegative: by the completeness of the instrument and the maximization definition of $f_{\cT}(\rho)=\max_{E\in\cT}\tr{E\rho}$, one obtains
$f_{\cT}(\rho)\le\sum_i p_i f_{\cT}(\rho_i)$,
which implies $\rD_{\cT}(\rho)\ge\sum_i p_i\rD_{\cT}(\rho_i)$ and guarantees $\Delta\rD_{\cT}(\{\Phi_i\})\ge0$. Introduce the Loewner downward target hull
\begin{equation}
 \mathord{\downarrow} \mathcal{C}=\{X=X^\dagger:X\leq\omega
 \text{ for some }\omega\in\cC\}.
\end{equation}
This set contains exactly the observables whose expectation is no larger than the target support function on every state.

We present the following formula as a spectral certificate proving the repairability of the instrument:
\begin{equation}
\mathfrak{S}_{\cT}(\{\Phi_i\}) \coloneqq \max_{(E_1,\ldots,E_n)\in\cT^n}
 \left[\min_{\omega\in\cC}
 \lamax\!\left(\sum_i\Phi_i^\dagger(E_i)-\omega\right)\right]_+.
\end{equation}
Here, $\lamax(\cdot)$ denotes the largest eigenvalue of a Hermitian operator, serving as the spectral gap that tests whether the pulled-back effect dominates the target reference. The operation $[\cdot]_+ = \max(0,\cdot)$ is the positive part, which acts as a safety truncation returning zero whenever the spectral gap is negative, thereby certifying that the instrument has no alignment margin.
This formula measures the maximal spectral alignment margin that a multi-outcome instrument can exert with respect to the target specification $\cT$. It is used as a key figure for analyzing not only the mere probabilistic success rate, but also whether and to what extent recovery is possible within the geometric structure of the state space.

\begin{thrm}
\label{thm:6}
Every finite instrument satisfies the exact identity
\begin{equation}
\Delta \rD_{\cT}(\{\Phi_i\}) = \mathfrak{S}_{\cT}(\{\Phi_i\}).
 \label{eq:14}
\end{equation}
Consequently, average deficiency is nondecreasing for every input,
\begin{equation}
 \sum_i p_i\rD_{\cT}(\rho_i)\geq \rD_{\cT}(\rho),
 \label{eq:15}
\end{equation}
if and only if every target tuple satisfies
\begin{equation}
 \sum_i\Phi_i^\dagger(E_i)\in\mathord{\downarrow} \mathcal{C}.
 \label{eq:16}
\end{equation}
In this case, the indicator $\mathfrak{S}_{\cT}(\{\Phi_i\})$ vanishes.
\end{thrm}

\begin{table*}[t]
\caption{One target-repair architecture instantiated in two settings. Each row follows the hierarchy: target orbit and pulled-back operator $\to$ matching certificate $\to$ operational boundary.}
\label{tab:commonarchitecture}
\centering
\includegraphics[width=0.99\textwidth]{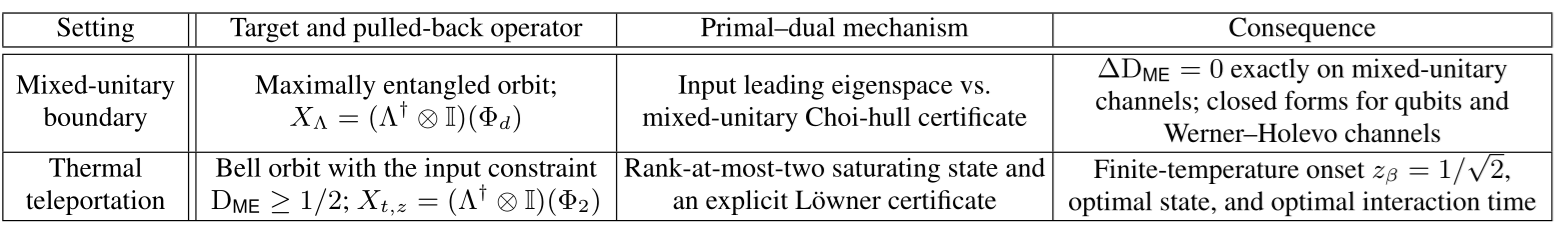}
\end{table*}

The underlying mechanism is simple yet powerful.
Our derivation is driven entirely by the task loss (deficiency) function, which transforms the affinity evaluation into a deficiency evaluation and introduces a minimization over $\omega \in \mathcal{C}$.
Applying Sion's minimax theorem exchanges the order of state and target-hull optimizations, reducing the nested optimizations directly to the spectral gap of the pulled-back operator $X_{\bm{E}} - \omega$.
This yields the spectral certificate $\mathfrak{S}_{\mathcal{T}}$.
Because this derivation is rooted in post-selection probabilities, $\mathfrak{S}_{\mathcal{T}} > 0$ if and only if $\Delta \mathcal{D}_{\mathcal{T}} > 0$, providing an absolute operational bound without requiring witness normalization.
(Detailed proofs are provided in Appendix~\ref{Appendix F}.)

The remainder of this section repeats the same proof architecture to establish clear operational boundaries for the two given cases.
Specifically, a target orbit is selected, and the input state is pulled toward this target orbit through concretely restricted dynamics.
Matching dualities and target-hull certificates are then cross-checked, and the operational boundary is explicitly derived.
Table~\ref{tab:commonarchitecture} makes these two corresponding cases explicit.

When the instrument reduces to a single deterministic channel, the general framework simplifies accordingly, leading to the following corollary.
For a deterministic channel $\Lambda$, this yields
\begin{equation}
 \Delta \rD_{\cT}(\Lambda)=\max_{\rho\in\D(\cH)}
 \bigl[\rD_{\cT}(\rho) - \rD_{\cT}(\Lambda(\rho))\bigr]\geq0,
\end{equation}
and the resulting consequences are summarized as follows.
The proof of the following corollary is given in the second part of Appendix~\ref{Appendix F}.

\begin{coro}
\label{cor:7}
For every channel $\Lambda$ and compact target family $\cT$,
\begin{equation}
\Delta \rD_{\cT}(\Lambda)=
 \max_{E\in\cT}\min_{\omega\in\cC}
 \lamax\!\bigl(\Lambda^\dagger(E)-\omega\bigr).
 \label{eq:18}
\end{equation}
If $\cT=\{E_1,\ldots,E_K\}$ is finite, each inner optimization is the SDP
\begin{equation}
 r_k=\min_{s,\bm q}\ s
 \quad\text{s.t.}\quad
 \Lambda^\dagger(E_k)-\sum_jq_jE_j\leq sI,
 \quad \bm q\in\Delta_K.
 \label{eq:19}
\end{equation}
Moreover, the following three inequalities hold:
\begin{align}
 |\Delta \rD_{\cT}(\Lambda)-\Delta \rD_{\cT}(\Gamma)|
 &\leq\half\|\Lambda-\Gamma\|_\diamond,
 \label{eq:20}\\
 \rD_{\cT}(\Lambda(\rho))
 &\geq \rD_{\cT}(\rho)-\Delta \rD_{\cT}(\Lambda).
 \label{eq:21}\\
 \Delta \rD_{\cT}(\Lambda_2\!\circ\!\Lambda_1)
 &\leq \Delta \rD_{\cT}(\Lambda_1)+\Delta \rD_{\cT}(\Lambda_2),
 \label{eq:22}
\end{align}
Consequently, from Eq.~\eqref{eq:20}, evaluating the bound for the nonrepairing set $\cN_{\cT}=\{\Gamma:\Delta \rD_{\cT}(\Gamma)=0\}$ yields the following upper bound:
\begin{equation}
 \Delta \rD_{\cT}(\Lambda)
 \leq\half\inf_{\Gamma\in\cN_{\cT}}
 \|\Lambda-\Gamma\|_\diamond.
 \label{eq:23}
\end{equation}
\end{coro}

The spectral minimizer in \eqref{eq:18} serves as a dual certificate on the target side, whereas a top-eigenspace state that saturates this bound acts as a primal witness. Their coexistence will be made explicit in both the Birkhoff and thermal scenarios.
Furthermore, the repair capacity defines an operational error budget. At the single-channel level, Eq.~\eqref{eq:21} dictates that the reduction in deficiency cannot exceed the channel's maximum capacity. Extending this to multistage experiments, Eq.~\eqref{eq:22} implies that the cumulative target improvement cannot exceed the sum of the individual capacities.
Finally, rather than being a mere uniform constant, Eq.~\eqref{eq:23} provides an individual geometric limit: the repair capacity $\Delta \rD_{\cT}(\Lambda)$ is bounded by half the diamond-distance from $\Lambda$ to the nearest nonrepairing channel. This implies that the repairability of a channel is fundamentally constrained by its structural separation from operations that are entirely incapable of target restoration.

\subsection{The mixed-unitarity bound for entanglement repairability}
\label{sec:4-a}

Consider the composite Hilbert space
$\cH_{AR} = \cH_A \otimes \cH_R \simeq \mathbb{C}^d \otimes \mathbb{C}^d$,
where $A$ and $R$ are uncorrelated reference systems of the same dimension $d$, and $R$ is left completely untouched by any channel.
Define the canonical maximally entangled (ME) state and its density operator as
\begin{equation}\label{24}
 |\Phi_d\rangle = \frac{1}{\sqrt{d}}\sum_{j=0}^{d-1}|j\rangle_A \otimes |j\rangle_R,
 \qquad \Phi_d = \proj{\Phi_d},
\end{equation}
where we adopt the normalized Choi representation for a quantum channel $\Lambda$:
\begin{equation}
 J(\Lambda) = (\Lambda_A \otimes \id_R)(\Phi_d).
\end{equation}
We consider the ME target orbit under local unitary rotations on $R$:
\begin{equation}
 \cT_{\textsf{ME}}^{(d)} = \bigl\{ (I_A \otimes U_R)\Phi_d(I_A \otimes U_R^\dagger) : U \in \mathsf{U}(d) \bigr\}.
 \label{eq:26}
\end{equation}
The deficiency with respect to this orbit is defined as $\rD_{\textsf{ME}}(\rho) = 1 - f_{\textsf{ME}}(\rho)$, which quantifies the distance from the maximally entangled target. Explicitly,
\begin{align}
 \rD_{\textsf{ME}}(\rho)
 &= \min_{\omega \in \cT_{\textsf{ME}}^{(d)}} \bigl(1 - \tr{\omega \rho}\bigr)\nonumber\\
 &= \min_{U \in \mathsf{U}(d)}\left\{1 -  \tr{(I \otimes U)\Phi_d(I \otimes U^\dagger)\rho}\right\},
\end{align}
where the minimum is taken over all states in the maximally entangled target orbit $\cT_{\textsf{ME}}^{(d)}$. The convex hull $\cM_d = \conv \cT_{\textsf{ME}}^{(d)}$ coincides precisely with the set of normalized Choi states of mixed-unitary channels ($\MU_d$).

We then pose the central dynamical question: can a local channel $\Lambda$ acting solely on subsystem $A$ repair the target entanglement? This is quantified by the repair capacity:
\begin{equation}
 \Delta \rD_{\textsf{ME}}(\Lambda) = \max_{\rho \in \D(\cH_{AR})}
 \bigl\{\rD_{\textsf{ME}}(\rho_{AR})- \rD_{\textsf{ME}}[(\Lambda_A \otimes \id_R)(\rho)]\bigr\}.
\end{equation}
Throughout this definition, the reference dimension is strictly fixed to $d$.

Prior work by Badzi\c{a}g \emph{et al.}~\cite{BadziagHorodeckiHorodeckiHorodecki2000} established that mixed-unitary dynamics cannot increase this score, proving the nonincrease theorem for unital channels in the qubit case ($d=2$). The converse result presented below identifies the missing higher-dimensional boundary. In this section, the term ``Birkhoff'' specifically denotes the finite-dimensional convex hull of mixed-unitary channels at a single channel use, distinct from the asymptotic quantum Birkhoff conjecture.

\begin{thrm}
\label{thm:8}
For every $d$-dimensional channel $\Lambda$, define the pulled Bell operator
\begin{equation}
 X_\Lambda=(\Lambda^\dagger\otimes\id)(\Phi_d).
\end{equation}
Then
\begin{equation}
 \Delta \rD_{\textsf{ME}}(\Lambda)=
 \min_{\omega\in\cM_d}\lamax(X_\Lambda-\omega),
 \label{eq:30}
\end{equation}
and
\begin{equation}
 \Delta \rD_{\textsf{ME}}(\Lambda)=0
 \quad\Longleftrightarrow\quad
 \Lambda\ \text{is mixed unitary}.
 \label{eq:31}
\end{equation}
Let
\begin{align}
 \delta_J(\Lambda)&=\inf_{\Omega\in\MU_d}
 \|J(\Lambda)-J(\Omega)\|_1,\nonumber\\
 \delta_\diamond(\Lambda)&=\inf_{\Omega\in\MU_d}
 \|\Lambda-\Omega\|_\diamond
\end{align}
denote the distances from $\Lambda$ to the mixed-unitary set in normalized Choi trace norm and diamond norm. Then
\begin{align}
 2\Delta \rD_{\textsf{ME}}(\Lambda)&\leq\delta_J(\Lambda)
 \leq2(d^2-1)\Delta \rD_{\textsf{ME}}(\Lambda),
 \label{eq:33}\\
 2\Delta \rD_{\textsf{ME}}(\Lambda)&\leq\delta_\diamond(\Lambda)
 \leq2d(d^2-1)\Delta \rD_{\textsf{ME}}(\Lambda).
 \label{eq:34}
\end{align}
\end{thrm}
\noindent (Appendix~\ref{Appendix G} supplies the full proof of \cref{thm:8}.)

The above statement is stronger than a one-sided witness:
non-mixed-unitarity is equivalent to the existence of a reference-assisted input state $\rho$ whose entanglement deficiency $\rD_{\textsf{ME}}$ strictly decreases under the local action of $\Lambda$.
In other words, the vanishing of the repair capacity $\Delta \rD_{\textsf{ME}}(\Lambda)=0$ provides a necessary and sufficient condition for membership in the mixed-unitary class.
Moreover, this increase quantitatively controls conventional distances to the mixed-unitary set in both directions, as shown in \eqref{eq:33} and \eqref{eq:34}.

The above statement is stronger than a one-sided witness: non-mixed-unitarity is equivalent to the existence of a
reference-assisted input state $\rho$ whose entanglement deficiency $\rD_{\textsf{ME}}$ strictly decreases under the local action
of $\Lambda$. In other words, the vanishing of the repair capacity $\Delta \rD_{\textsf{ME}}(\Lambda)=0$ provides a necessary and
sufficient condition for membership in the mixed-unitary class. Moreover, this increase quantitatively controls conventional distances
to the mixed-unitary set in both directions, as shown in \eqref{eq:33} and \eqref{eq:34}. This demonstrates that $\Delta \rD_{\textsf{ME}}$
reflects the absolute change in the entanglement-assisted pass probability and admits a more direct operational interpretation.

\subsubsection{Qubit case: quantifying the birkhoff gap via entanglement deficiency}
\label{subsec:qubit}

For qubits ($d=2$), the quantum Birkhoff theorem guarantees that every unital channel is mixed unitary \cite{MendlWolf2009}.
While this is a qualitative classification, it leaves the quantitative gap---how far a non-unital channel deviates from this set---unmeasured. Our target-repair framework bridges this gap by deriving a closed-form formula for the repair capacity, $\Delta \rD_{\textsf{ME}}(\Lambda)$. This quantity determines the maximal reducible deficiency across all qubit channels, capturing the exact deviation from the unital (mixed-unitary) set.

To state the result, we briefly recall the affine Bloch representation of a qubit channel $\Lambda$.
Any single-qubit density operator can be expanded as $\rho = \frac{1}{2}(I + \bm r \cdot \bm \sigma)$, where $\bm r = (r_x, r_y, r_z)^{\mathsf T} \in \mathbb{R}^3$ is the Bloch vector with $\|\bm r\|_2 \le 1$, and $\bm \sigma = (\sigma_x, \sigma_y, \sigma_z)^{\mathsf T}$ is the vector of Pauli matrices.
A general qubit channel $\Lambda$ maps Bloch vectors according to an affine transformation:
\begin{equation}
 \bm r \;\longmapsto\; M\bm r + \bm t.
\end{equation}
Here, $M \in \mathbb{R}^{3\times 3}$ is a real matrix governing the linear (rotational and contracting) part of the dynamics, while $\bm t \in \mathbb{R}^3$ is a translation vector.
Complete positivity and trace preservation constrain $M$ and $\bm t$ such that the image of the Bloch sphere remains within the valid Bloch ball.
Crucially, the channel is \emph{unital} ($\Lambda(I)=I$) if and only if $\bm t = \bm 0$; thus, $\bm t$ quantifies the nonunital part of $\Lambda$.

\begin{thrm}
\label{thm:9}
Let a qubit channel have affine Bloch representation
$\bm r\mapsto M\bm r+\bm t$. Then
\begin{equation}
 \Delta \rD_{\textsf{ME}}(\Lambda)=\frac{\|\bm t\|_2}{4}
 =\frac14\|\Lambda(I/2)-I/2\|_1.
 \label{eq:39}
\end{equation}
An optimal input can always be chosen separable, in the form
$I/2\otimes\proj{\psi_{\bm t}}$.
\end{thrm}
\noindent The proof of this result is given in Appendix~\ref{Appendix H}.

This result illustrates how the high-dimensional spectral duality (Theorem~\ref{thm:8}) collapses into a simple Bloch-vector formula in the qubit regime.
Crucially, it reveals that maximal-entanglement repair measures precisely the nonunital part (the translation $\bm t$) of qubit dynamics, even though the optimizer achieving this repair need not itself be entangled.
In other words, while the target is maximally entangled, the \emph{input} can be chosen to be completely separable.

The unital nonincrease part is known \cite{BadziagHorodeckiHorodeckiHorodecki2000}, and correlation-matrix bounds for local nonunital channels were recently developed in Ref.~\cite{LiuLi2025}.
The advance in \cref{thm:9} is the exact capacity equality for all qubit channels, the identification of a separable state that saturates this bound, and its consistent embedding within the all-dimensional mixed-unitary characterization established earlier.

It should be emphasized that the reduction of deficiency by a local non-unital channel does not violate entanglement monotonicity.
As noted earlier, this reduction is consistent with the fact that entanglement deficiency does not generally satisfy universal monotonicity under local quantum operations (see~\cref{Violation of universal monotonicity in entanglement}).
A crucial point to note is that the deficiency $\rD_{\textsf{ME}}(\rho)$ is evaluated solely relative to the best-aligned point within the local-unitary orbit of maximally entangled states.
Consequently, when a non-unital channel shifts the local Bloch vector by $\bm t\neq \bm 0$, it realigns the existing entanglement to match one particular member of this unitary orbit.
As a result, the deficiency decreases even though the actual amount of entanglement does not increase.
Thus, a positive repair capacity $\Delta \rD_{\textsf{ME}}>0$ certifies a structural realignment of existing entanglement rather than a creation of new entanglement.

\subsubsection{Higher dimensions: Positive repair capacity in unital Werner--Holevo channels}
\label{subsec:WH}

\begin{figure*}[t]
\centering
\includegraphics[width=0.91\textwidth]{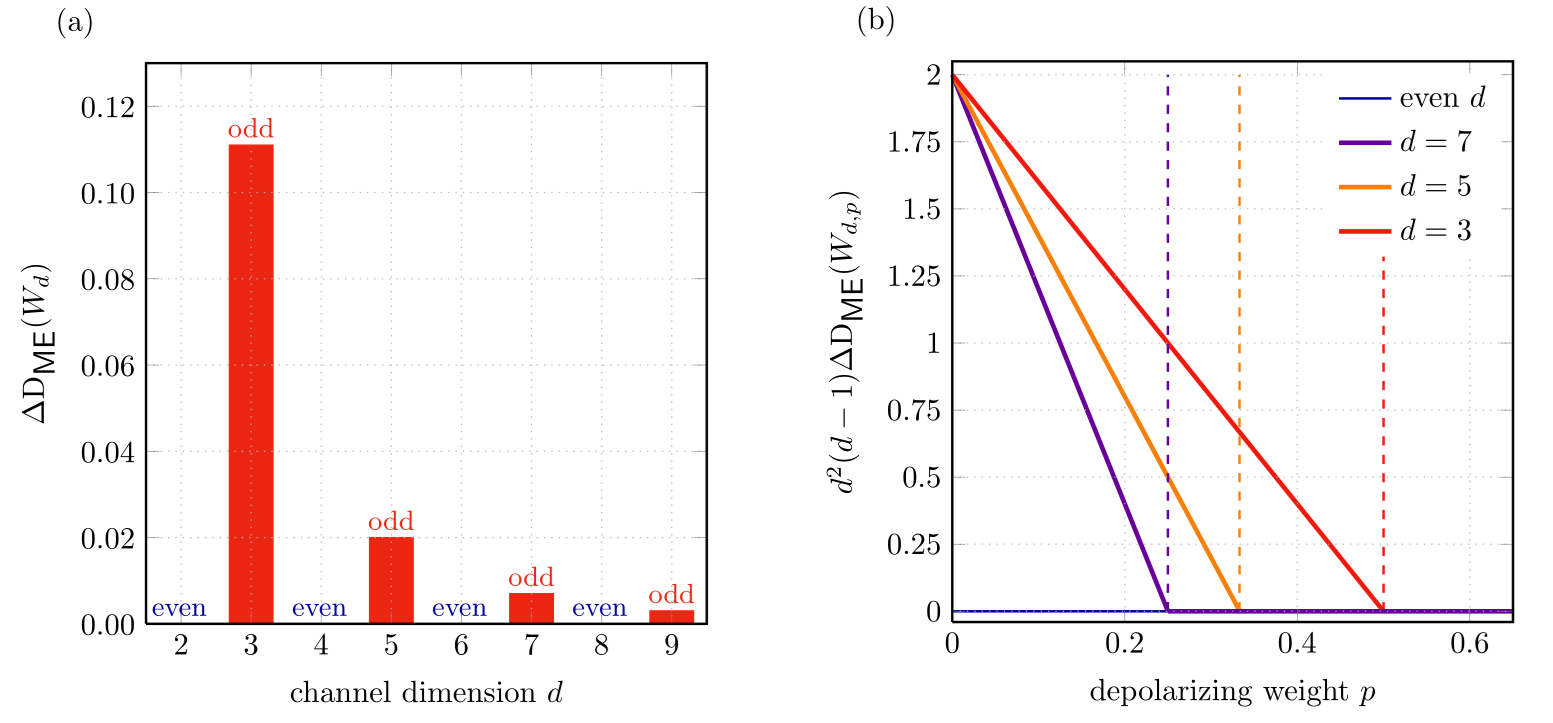}
\caption{The repair capacity provided by the Werner--Holevo channels.
(a) The unital channel $W_d$ exhibits zero repair capacity in even dimensions, whereas in odd dimensions it takes a strictly positive value equal to $2/[d^2(d-1)]$.
(b) For odd $d$, the dimension-rescaled repair decays linearly under depolarizing noise and vanishes at the threshold $p = 2/(d+1)$; in contrast, every even-dimensional curve is identically zero.}
\label{fig:birkhoff}
\end{figure*}

The qubit formula \eqref{eq:39} might suggest that only the nonunital (translational) part of a channel can repair maximally entangled targets.
Higher dimensions disprove this interpretation.
To demonstrate this, consider the unital Werner--Holevo channel and its depolarized version, a noise model related to previously studied noisy Werner--Holevo families \cite{Karimipour2024},
\begin{align}
 W_d(\rho)&=\frac{\tr{\rho}I-\rho^{\mathsf T}}{d-1},
\qquad  \Delta_d(\rho)=\tr{\rho}\frac I d,\nonumber\\
 W_{d,p}&=(1-p)W_d+p\Delta_d.
\end{align}
Both $W_d$ and the depolarized channel $W_{d,p}$ are unital, i.e.\ $\Lambda(I/d)=I/d$.
Let $F$ denote the swap operator, and let $P_\pm=(I\pm F)/2$ be the symmetric and antisymmetric projectors and $r_\pm=d(d\pm1)/2$ their ranks.

\begin{thrm}
\label{thm:10}
For $0\leq p\leq1$,
\begin{equation}
 \Delta \rD_{\textsf{ME}}(W_{d,p})=
 \begin{cases}
 0,&d\ \text{even},\\[1mm]
 \displaystyle\frac{[2-(d+1)p]_+}{d^2(d-1)},&d\ \text{odd}.
 \end{cases}
 \label{eq:41}
\end{equation}
For odd $d$ and $p<2/(d+1)$, the normalized antisymmetric state
$\rho_*=P_-/r_-$ is a primal optimizer. Transpose-depolarizing noise removes repair exactly at $p_{\rm MU}=2/(d+1)$.
\end{thrm}
\noindent (Appendix~\ref{Appendix I} provides the proof of the above theorem.)

This result promotes the known mixed-unitary parity of $W_d$ \cite{MendlWolf2009,GirardLeungLevickLiPaulsenPoonWatrous2022} to an exact repair law with an explicit noise threshold $p_{\rm MU}$.
Crucially, it shows that unital channels ($\bm t=0$) can possess strictly positive repair capacity in odd dimensions, thereby falsifying the qubit-inspired intuition that only nonunital translations matter. Figure~\ref{fig:birkhoff}(a) shows the repair capacity provided by the Werner-Holevo channels across different dimensions, while Fig.~\ref{fig:birkhoff}(b) illustrates the variation of the repair capacity with respect to depolarizing noise in several dimensions.

The qutrit case ($d=3$) makes both sides of the duality concrete.
Recall that the entanglement deficiency is defined as $\rD_{\textsf{ME}}(\rho) = 1 - f_{\textsf{ME}}(\rho)$, measuring the minimal deviation from the maximally entangled target orbit $\cT_{\textsf{ME}}^{(d)}$.
For the Werner--Holevo channel $W_3$, we have
\begin{equation}
 \Delta \rD_{\textsf{ME}}(W_3)=\frac19,\quad
 \rho_*=\frac{P_-}{3},\quad
 \omega_*=\frac29P_-+\frac1{18}P_+.
\end{equation}
Here, the input deficiency is $\rD_{\textsf{ME}}(\rho_*) = 7/9$, while the output deficiency is $\rD_{\textsf{ME}}[(W_3\otimes\id)(\rho_*)] = 2/3$.
Consistent with the duality theorem, the reduction in deficiency $7/9 - 2/3 = 1/9$ exactly matches the spectral gap $\lambda_{\max}(P_-/3-\omega_*) = 1/9$.

Crucially, the output score $1/3$ coincides precisely with the qutrit classical teleportation boundary.
One must not interpret this as implying that some unital channel can surpass this classical threshold in teleportation-like tasks.
Since our output reaches but does not exceed this classical threshold, one must not overstate this result as a breakthrough in qutrit teleportation.
Rather, the result serves strictly as a witness of non-random-unitary repair: it rigorously certifies that $W_3$ is not a mixed-unitary channel, without implying any quantum advantage in transmission fidelity. However, we will show in what follows that, for certain non-unital channels, surpassing this classical threshold is indeed possible, which we establish via an explicit closed-form formula.

\subsection{Global thermal activation of teleportation via deficiency optimization}
\label{sec:4-b}
For a two-qubit state $\rho$, the deficiency is defined as
\begin{equation}
 \rD_{\textsf{ME}}(\rho) =\min_{U\in\mathsf U(2)}\bigl[1-\langle\Phi_2|(I\otimes U)\rho(I\otimes U^\dagger)|\Phi_2\rangle\bigr],
\end{equation}
which quantifies the distance from $\rho$ to the target orbit $\cT_{\textsf{ME}}^{(2)}$ of the maximally entangled state $|\Phi_2\rangle$.
The standard teleportation fidelity is $(2f_{\textsf{ME}}+1)/3$, so a state outperforms the best classical protocol exactly when $\rD_{\textsf{ME}}(\rho)<1/2$ \cite{Bennett2, Horodecki3}.
The target set $\cT_{\textsf{ME}}^{(2)}$ is therefore the two-qubit version of \eqref{eq:26}.

The enhancement of teleportation fidelity (i.e., the reduction of $\rD_{\textsf{ME}}$) by local amplitude damping was first exhibited for special input families \cite{BadziagHorodeckiHorodeckiHorodecki2000, Bandyopadhyay2002}, and related local-noise settings were subsequently explored \cite{Yeo2008}.
The question solved here is stronger: optimize over every input below the classical deficiency boundary ($\rD_{\textsf{ME}}>1/2$) and over the interaction time, at arbitrary bath polarization, and certify the answer by an explicit operator inequality.

Generalized amplitude damping (GAD) on the first qubit has affine Bloch action
\begin{equation}
 \bm r\longmapsto
 A\bm r+\gamma z_\beta\bm e_z,\ \
 A=\diag(\sqrt{1-\gamma},\sqrt{1-\gamma},1-\gamma),
\end{equation}
where $z_\beta=\tanh(\beta E/2)$ is the bath polarization with $Z|0\rangle=|0\rangle$, and $\bm e_z=(0,0,1)^T$ denotes the thermal equilibrium axis toward which the state is attracted, whose attraction strength is determined by the energy gap $E$.
The GAD channel models the finite-temperature
dissipation of a qubit coupled to a thermal bath; unlike zero-temperature
amplitude damping, it allows both relaxation and thermal excitation, and its
unique stationary state is the thermal (Gibbs) state
$\rho_{\rm th}=\tfrac12(I+z_\beta\sigma_z)$ \cite{Nielsen}.
Crucially, because $\gamma z_\beta\bm e_z\neq\bm0$, this channel is non-unital,
possessing a non-zero translation vector $\bm t$.
Physically, this translation originates from the energy exchange between the
system and a thermal bath at inverse temperature $\beta$, driving the state
towards the thermal equilibrium of the bath.
At fixed temperature, a Markov realization can be written as
$\gamma(\tau)=1-e^{-\Gamma\tau}$, so varying $\gamma$ corresponds to varying
the interaction time rather than changing the bath.

The unconstrained repair problem is now an immediate consequence of \cref{thm:9}.
Specifically, the repair capacity is given by the norm of the translation vector,
\begin{equation}
 \Delta \rD_{\textsf{ME}}(\Lambda_{\gamma,z_\beta})=\frac{\|\bm t\|_2}{4}=\frac{\gamma z_\beta}{4}>0,
\end{equation}
which explicitly certifies that the GAD channel can reduce the deficiency $\rD_{\textsf{ME}}$ for suitably chosen inputs without ever increasing the underlying entanglement.

By \cref{thm:9}, the optimal input can always be chosen as a separable state of the form $\rho=I/2\otimes\proj{\psi_{\bm t}}$, where $\psi_{\bm t}$ is a pure state whose Bloch vector points along the direction of $\bm t=\gamma z_\beta\bm e_z$.
For such an input, the output deficiency satisfies
\begin{equation}
 \rD_{\textsf{ME}}(\Lambda_{\gamma,z_\beta}\otimes\id(\rho))
 = \rD_{\textsf{ME}}(\rho)-\frac{\gamma z_\beta}{4}.
\end{equation}
Therefore, the classical teleportation boundary $D_{\rm out}<1/2$ is surpassed whenever
\begin{equation}
\rD_{\textsf{ME}}(\rho)-\frac{\gamma z_\beta}{4}<\frac12,
 \qquad
 \rD_{\textsf{ME}}(\rho)>\frac12,
\end{equation}
or equivalently $\gamma z_\beta>2(2\rD_{\textsf{ME}}(\rho)-1)$.

In the Choi matrix language, the above certification is equivalent to the operator inequality
\begin{equation}
 X_{\Lambda_{\gamma,z_\beta}}
 > \frac14 P_+ + \Bigl(\frac12-\frac{|\gamma z_\beta|}{4}\Bigr)P_-,
 \label{eq:48}
\end{equation}
where $P_\pm$ are the symmetric and antisymmetric maximally entangled projectors, respectively.
To see this, recall that for a single-qubit channel the Choi matrix $X_\Lambda$ is diagonal in the Bell basis.
For the GAD channel $\Lambda_{\gamma,z_\beta}$ with affine parameters $(A,\bm t)$, the eigenvalue in the symmetric sector is
$\lambda_+(X_{\Lambda_{\gamma,z_\beta}}) = (2-\gamma+2\gamma z_\beta)/4$,
while the minimal eigenvalue in the antisymmetric sector is
$\lambda_{\min}^-(X_{\Lambda_{\gamma,z_\beta}}) = (2-\gamma-2\gamma z_\beta)/4$.

The condition for surpassing the classical boundary, $\gamma z_\beta > 2(2\rD_{\textsf{ME}}(\rho)-1)$, implies that the antisymmetric component of the output state is sufficiently suppressed.
Consequently, the minimal eigenvalue in the antisymmetric sector, $\lambda_{\min}^-$, becomes strictly larger than the classical threshold $1/2 - \gamma z_\beta/4$, yielding \eqref{eq:48}.

The strict positivity of the spectral gap in \eqref{eq:48} follows directly from the spectral analysis of $X_{\Lambda_{\gamma,z_\beta}}$.
Physically, the non-unital displacement $\bm t$ shifts the center of mass in the $P_-$ subspace.
As $\gamma z_\beta$ increases, it elevates the minimal eigenvalue of $P_-$ from $1/4$ to $1/2 - \gamma z_\beta/4$.
This signifies that the channel acts to compress the antisymmetric sector, thereby preserving and enhancing entanglement.
Thus, \eqref{eq:48} provides an explicit operator inequality demonstrating the possibility of crossing the classical boundary without any numerical optimization.

This finding stands in sharp contrast to earlier investigations \cite{BadziagHorodeckiHorodeckiHorodecki2000,Bandyopadhyay2002,Yeo2008}, which reported fidelity enhancements confined strictly to specialized input ensembles or highly specific noise models.
In contrast, by leveraging \cref{thm:9}, we establish the optimal input state, the maximal attainable repair, and the exact threshold for surpassing the classical teleportation boundary, all derived in closed analytical form for arbitrary interaction times $\gamma$ and bath polarizations $z_\beta$.
Furthermore, framing the entire analysis in terms of the entanglement deficiency $\rD_{\textsf{ME}}$ reveals a profound spectral target-repair duality.
This duality yields a transparent geometric interpretation within Choi space: a channel possesses a strictly positive repair capacity if and only if its Choi matrix $X_\Lambda$ resides strictly outside the downward closure $\mathord{\downarrow}\cM_2$ of the mixed-unitary set $\cM_2$ (i.e., $X_\Lambda\notin\mathord{\downarrow}\cM_2$).
Consequently, the non-unital displacement vector $\bm t$ of the GAD channel emerges as the sole geometric invariant governing the repairability of entanglement deficiency.
This not only completes the deficiency-driven narrative of this work but also elevates the Choi-matrix perspective to a central diagnostic tool for quantum thermometry and resource manipulation.

This capacity indicates how much the deficiency can decrease, but it does not directly show whether an initially nonuseful state can cross the classical threshold.
To clarify our objective, we thus present the following two definitions.
The minimum deficiency frontier is given by
\begin{equation}
 \mathcal{D}(t,z)=
 \min_{\rho:\,\mathcal{D}_{\textsf{ME}}(\rho)\geq\frac12}
  \mathcal{D}_{t,z}^{\mathrm{out}}(\rho),
 \quad t=\sqrt{1-\gamma},
\end{equation}
where $\mathcal{D}_{t,z}^{\mathrm{out}}(\rho) = \mathcal{D}_{\textsf{ME}}[(\Lambda_{1-t^2,z}\otimes\id)(\rho)]$.
Physically, $\mathcal{D}(t,z)$ represents the optimal reduced deficiency (as a result of repair) at fixed time and temperature.
And then the global activation above the classical boundary is given by
\begin{equation}
 \mathcal{A}(z)=\max_{0\leq t\leq1}\bigl[\frac{1}{2} - \mathcal{D}(t,z)\bigr].
\end{equation}
Here, $\mathcal{A}(z)$ quantifies the globally optimal amount of activation surpassing the classical boundary at a given temperature.
Because $t=1$ ($\Longleftrightarrow \gamma=\tau=0$) is the identity evolution, $\mathcal A(z)\geq0$.
To derive the activation threshold and the global optimal state for $\mathcal{A}(z)$, we first present the closed-form expression for the deficiency frontier $\mathcal{D}(t,z)$, the critical threshold for $0<t<1$, and the global optimal state from the following lemma.

\begin{lem}
\label{lem:11}
For every $t,z\in[0,1]$,
\begin{equation}
 \mathcal D(t,z))=\frac{3-\ell(t,z)}{4},
\end{equation}
where
\begin{equation}
 \ell(t,z)=t+(1-t^2)z
 \frac{t+(1+t)z}{2t+(1+t)z},
\end{equation}
with the continuous value $\ell(0,0)=0$. An optimal input is the rank-at-most-two state
\begin{equation}
 \rho_u=\frac14
 \begin{pmatrix}
 (1+u)^2&0&0&1-u^2\\
 0&0&0&0\\
 0&0&2(1-u^2)&0\\
 1-u^2&0&0&(1-u)^2
 \end{pmatrix},
 \label{eq:53}
\end{equation}
where
\begin{equation}
 u=\frac{(1+t)z}{2t+(1+t)z}.
 \label{eq:54}
\end{equation}
At the removable point $(t,z)=(0,0)$ we set $u=0$. In every case
$\rD_{\textsf{ME}}(\rho_u)=1/2$ exactly. For $0<t<1$, strict activation at the prescribed interaction time occurs if and only if
\begin{equation}
 z>z_c(t):=
 \frac{1-t+\sqrt{1+6t+t^2}}{2(1+t)}.
 \label{eq:55}
\end{equation}
\end{lem}

The global nature of this result is important. The derivation may be visualized by twirling with $R_z(\theta)\otimes R_z(-\theta)$, which leaves the pulled Bell test invariant and cannot violate the constraint $\rD_{\textsf{ME}}\geq1/2$. This twirl reduces the input to a real $X$ state. Positivity and the threshold constraint then specify a one-parameter physical boundary whose objective is a concave quadratic---strictly concave away from the degenerate boundaries---and at which the vertex yields \eqref{eq:53}. A detailed proof of this derivation is provided in Appendix~\ref{Appendix J}.

As shown in the preceding lemma, optimizing the interaction time yields a genuine phase boundary.
Building on this, the following theorem establishes the threshold and the global optimal state for activation within $\mathcal{A}(z)$.
Physically, this implies that the transition from thermal degradation to strict activation constitutes a sharp, well-defined phase transition governed entirely by the interplay between the interaction time and the thermal parameter $z$. A detailed proof is provided in Appendix~\ref{Appendix K}.

\begin{thrm}
\label{thrm:12}
The non-surpassing activation obeys
\begin{equation}
 \mathcal A(z)=0\quad\Longleftrightarrow\quad z\leq\frac1{\sqrt2}.
\end{equation}
For $z>1/\sqrt2$, then
\begin{equation}
 \mathcal A(z)= \frac{(1-t_*(z))\big[K^2(z)-2\big]}{8}>0,
\end{equation}
where $K(z)=(1+t_*(z))z$, and $$t_*(z)=\frac{1-z^2+\sqrt{1+2z^2}}{z(2+z)}$$ is the optimal interaction variable.
The optimal input is
\begin{equation}
 \rho_*(z) = \frac{1}{4}
 \begin{pmatrix}
 (1+u_*(z))^2 & 0 & 0 & 1-u_*^2(z) \\
 0 & 0 & 0 & 0 \\
 0 & 0 & 2(1-u_*^2(z)) & 0 \\
 1-u_*^2(z) & 0 & 0 & (1-u_*(z))^2
 \end{pmatrix},
\end{equation}
where $u_*(z)=\frac{\sqrt{1+2z^2}+z-1}{2+z}$ (or equivalently, $K(z) - 1$).
\end{thrm}

\begin{figure*}[t]
\centering
\includegraphics[width=0.93\textwidth]{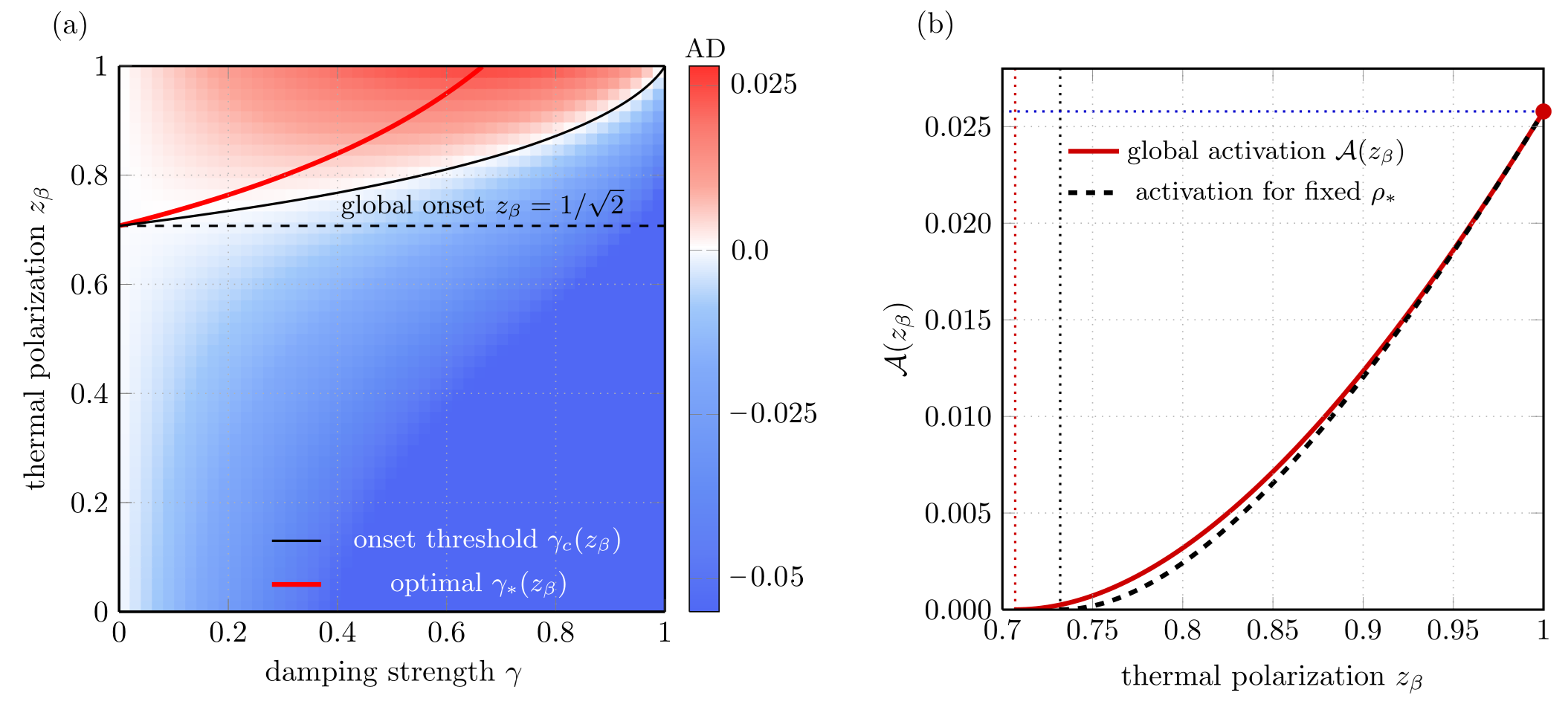}
\caption{Global thermal activation. (a) The exact activation deviation (AD), $1/2 - \mathcal{D}(\sqrt{1-\gamma}, z_\beta)$, at fixed damping and polarization. The black curve is the analytic activation critical line given by $t_c(z_\beta)$; the red curve is the globally optimal interaction trajectory given by $t_*(z_\beta)$. (b) The global activation $\mathcal A(z_\beta)$, and the fixed-$\rho_*$ result. The global activation onset $1/\sqrt{2}$ (red dotted line) precedes the activation threshold $\sqrt{3}-1$ (black dotted line) for the fixed input $\rho_*$, and both yield the same optimal activation value of $(2\sqrt{3}-3)/18$ (blue dotted line) at zero temperature.}
\label{fig:globalthermal}
\end{figure*}

The threshold in the above theorem corresponds to
\begin{equation}
 \beta E>2\operatorname{artanh}(1/\sqrt2)
 =2\ln(1+\sqrt2)\simeq1.762747,
\end{equation}
and, for $z>1/\sqrt2$, activation occupies a finite ``Goldilocks'' interval
\begin{equation}
 t_c(z)<t<1,\qquad
 t_c(z)=\frac{1-z^2+\sqrt{1-z^2}}{z(1+z)}.
\end{equation}
Consequently, the activation phenomenon ceases to exist whenever $t_c(z) \ge 1$ ($\Longleftrightarrow z\leq 1/\sqrt2$).
Equivalently, $0<\gamma<1-t_c^2$. Neither zero interaction nor complete thermalization gives a strict crossing. In the Markov parameterization, the optimal time is
$\tau_*=-2\Gamma^{-1}\ln t_*$. Derived from these results, Fig.~\ref{fig:globalthermal}(a) provides the global thermal activation profile at fixed damping and polarization, along with the activation critical line and the globally optimal interaction trajectory.

At zero temperature, $z=1$, one has $t_*(1)=u_*(1)=1/\sqrt3$ and $\gamma_*=2/3$. The optimal state becomes
\begin{equation}
 \rho_*=\frac23\proj{\phi_q}+\frac13\proj{10},
 \qquad q=\frac{2+\sqrt3}{4},
\end{equation}
where $|\phi_q\rangle=\sqrt q|00\rangle+\sqrt{1-q}|11\rangle$. The optimized figure of deficiency is
\begin{equation}
 \mathcal D(1/\sqrt3,1)=\frac{6-\sqrt3}{9}.
\end{equation}
The corresponding largest activation gain is
\begin{equation}
 \mathcal A(1)=\frac{2\sqrt3-3}{18}\simeq0.025783.
\end{equation}
The earlier explicit construction of Ref.~\cite{BadziagHorodeckiHorodeckiHorodecki2000} reached approximately $1-0.5224$ in output deficiency; the present work establishes that the optimal deficiency is $1 - 0.52578\ldots$ and extends this result to finite temperatures.
Consequently, the state previously linked to the threshold $z>\sqrt{3}-1$ is not invalidated but rather identified as the strict zero-temperature endpoint ($z=1$) of the optimal state family $\rho_u$.
Crucially, away from absolute zero, this state can no longer remain fixed as the optimal state: it fails to exploit thermal fluctuations and misses the strict activation window $1/\sqrt{2}<z\leq\sqrt{3}-1$, whereas the dynamically optimized family $\rho_u$ yields a strictly lower output deficiency in this regime.

 Thus the state previously associated with the threshold $z>\sqrt3-1$ is not discarded: it is the zero-temperature endpoint of the optimal family. Away from $z=1$, keeping this state fixed is suboptimal and misses the activation window
$1/\sqrt2<z\leq\sqrt3-1$. A comparison between the optimal activation trajectory for the fixed input $\rho_*$ and the globally optimal activation trajectory is illustrated in Fig.~\ref{fig:globalthermal}(b).

It should be emphasized that such activation does not constitute entanglement generation.
The input concurrence for the state in Eq.~\eqref{eq:53} is $(1-u^2)/2$, and a local channel cannot increase this value.
In the zero-temperature case, even though the teleportation fidelity surpasses the classical limit, the concurrence drops from $1/3$ to $(\sqrt{3}-2\sqrt{2-\sqrt{3}})/9$.
Specifically, the GAD channel consumes entanglement while repairing the alignment with the maximally entangled target orbit.
While these phenomena have been noted in passing in several prior works, they have not been subjected to a systematic analysis within a concrete theoretical framework.
Our target-oriented framework provides a rigorous formalism to expose and systematize this essential distinction.

\section{Conclusion}
\label{sec:conclusion}

Several recent studies have highlighted operational scenarios that are ill-suited for conventional quantum resource theories (QRTs).
The task of entanglement-assisted teleportation epitomizes such situations; in other goal-oriented quantum protocols as well, the failure of the distance from free states to characterize quantum efficiency renders the definition of free state sets and free operations ambiguous, thereby diminishing the operational significance of ``free'' states in practical quantum tasks.
Motivated by this, we abandoned the notion of ``free'' resources and established a goal-oriented framework to quantitatively assess quantum efficiency---a necessary shift to unify previous piecemeal, task-specific approaches lacking a formal structure.

Our framework quantifies the deficiency relative to the maximal resource (efficiency) set.
We demonstrated that the resulting geometric deficiency measure rigorously characterizes the operational weakness in sub-channel quantum discrimination and proves highly suitable for quantifying quantum channel noise.
Furthermore, to demonstrate how the deficiency measure functions as a metric for mission success when the role and definition of the free set are ambiguous---particularly given a quantum target specification---we defined an operational quantity called ``repair capacity.''
We proved that this capacity equals a one-sided spectral separation of the pulled-back effects from the target hull, thereby exposing the mixed-unitary channel boundary regarding the repairability of multi-dimensional quantum entanglement.

Building on this theoretical rigor, we derived the affine Bloch translation law for qubit channels and the odd-even separation for Werner--Holevo channels.
Specifically, for one-sided qubit generalized amplitude damping (GAD) channels, optimizing over all inputs below the classical teleportation limit yields a closed-form relation for global activation, revealing the finite-temperature onset $z_\beta = 1/\sqrt{2}$ and the optimal interaction time.
In this context, we confirm that the global onset $1/\sqrt{2}$ precedes the threshold $\sqrt{3}-1$ for the fixed input state $\rho_*$, which yields the optimal activation value $(2\sqrt{3}-3)/18$.
This analytically demonstrates, with explicit numerical values, that global optimization can shift the critical parameter to a lower value compared to the local threshold, even for states possessing optimal activation values.

These results clearly reveal the central role played by the purpose-oriented framework in practical quantum tasks. Without
presupposing a predefined free-state manifold, our framework enables a direct assessment of a channel's utility and task feasibility
solely from the user's target specification. In such purpose-oriented settings, the deficiency measure serves as an operational indicator
that quantitatively bridges the gap between ideal goals and realistic constraints, whether the dynamics are non-random-unitary, algorithmic
calibration is required, or task activation in a dissipative environment is at stake.

Moreover, beyond the target affinity and its associated deficiency studied here, we anticipate that deficiency measures induced by
general distance measures can be extended to diverse quantum missions. This demonstrates that the purpose-oriented framework transcends a
mere mathematical alternative and offers a practical, integrative paradigm directly applicable to quantum technology design and benchmarking.

{\it Acknowledgments.--}This work is supported by the Fundamental Research Funds for the Central Universities (Grants No. 3072025YC2404), National Natural Science Foundation of China (Grants No. 12201555), Natural Science Foundation of Hunan province (Grants No. 2025JJ50050) and Hunan Basic Science Research Center for Mathematical Analysis (2024JC2002).

\begin{widetext}

\appendix

\section{Proofs of Theorems 1 and 2}
\label{Appendix A}

The geometric function $\rD_g$ we defined in main text always holds the conditions of (D1) and (D2) for any quantum resource.
First, we check about (D1). The fidelity between all quantum states is less than or equal to $1$, so $\rD_g(\rho)\geq 0$. And if $\rho$ is in the maximum resource state, from that definition, $\rD_g(\rho)= 0$, on the contrary, if $\rD_g(\rho)= 0$, there exists $\sigma$ that satisfies $\rF(\rho,\sigma)=1$, which means $\rho=\sigma$, where $\rho$ is the maximum resource state.
After this, for convenience, we define the maximum fidelity under the maximum resource states as follows:
\be
\rF_{\cR}(\rho) = \max_{\sigma\in\cS^{\max}}\rF(\sigma,\rho)
\ee
Next, from the convexity of the maximum value function $\rF_{\cR}$,
\begin{eqnarray}\label{7}
\rD_g(\sum_iq_i\sigma_i) = 1-\rF_{\cR}(\sum_iq_i\sigma_i) \geq 1-\sum_iq_i\rF_{\cR}(\sigma_i) = \sum_iq_i\rD_g(\sigma_i),
\end{eqnarray}
therefore, (D2) is satisfied.

\subsection{Proof of Theorem~\ref{thm1}}
We only need to prove that it satisfies (D3b) to confirm that $\rD^C_g$ is the measure of deficiency. By the arbitrary maximally coherent states $\out{\psi}{\psi}$ are in the form of $\ket{\psi} = \frac{1}{\sqrt{d}}\sum_{i}e^{i\theta_i}\ket{i}$, they are derived as follows:
\be
\rF_{\cC}(\rho) = \max_{\sigma\in\cC^{\max}}\rF(\sigma,\rho)= \frac{1}{d}\max_{\{\theta_i\}_i}\{\sum_{i,j}e^{i\theta_{ij}}\rho_{ij}\}
\ee
where $\rho_{ij} = \iinner{i}{\rho|j}$ and $e^{i\theta_{ij}} = e^{i(\theta_j-\theta_i)}$. It implies that $\rF_{\cC}(\rho) \leq \frac{1}{d}\sum_{i,j}|\rho_{ij}|$, and the equation holds if $\rho$ is pure.
Therefore, for quantum coherence, we have
\be\label{1-1}
\rD^C_g(\rho) \geq 1-\frac{\sum_{i,j}|\rho_{ij}|}{d}
\ee
and if $\rho$ is pure, we have
\be\label{1-2}
\rD^C_g(\rho) = 1-\frac{\sum_{i,j}|\rho_{ij}|}{d}.
\ee

Let $\Phi_\pi$ is an incoherent operation acted on by a series of permutation matrices $\{P_{\pi_n}\}$, \emph{e.i.}, $\Phi_\pi(\rho) = \sum_np_nP_{\pi_n}\rho P_{\pi_n}^\dagger$.
Then, there is a set $\{\theta^{(n)}_i\}_i$ that reach the maximum of $\rF_{\cC}(\rho_n)$ with $\rho_n = P_{\pi_n}\rho P_{\pi_n}^\dagger$, it implies that
\begin{eqnarray}\label{1-3}
\sum_np_n\rF_{\cC}(\rho_n) = \sum_np_n\sum_{i,j}e^{i\theta^{(n)}_{ij}} \rho_{\pi^{-1}_n(i)\pi^{-1}_n(j)}
\leq  \sum_{i,j}e^{i\theta_{ij}}\rho_{ij} =\rF_{\cC}(\rho)
\end{eqnarray}
because $\sum_{i,j}e^{i\theta^{(n)}_{ij}} \rho_{\pi^{-1}_n(i)\pi^{-1}_n(j)}\leq \sum_{i,j}e^{i\theta_{ij}}\rho_{ij}$ for any $n$.
Hereby, we have that $\rD^C_g(\rho) \leq \sum_np_n\rD^C_g(P_{\pi_n}\rho P_{\pi_n}^\dagger).$

Next, for any incoherent operation $\Phi$, acting as $\Phi(\rho) = \sum_n K_n\rho K_n^\dagger$, let $k^{(n)}_j\ (j = 1,2,\cdots,d)$ be the nonzero element at the $j$th column of $K_n$ (if there is no nonzero element in the $j$th column, then $k^{(n)}_j=0$). Suppose $k^{(n)}_j$ locates the $f_n(j)$th row. Here, $f_n(j)$ is a function that maps $\{2,\cdots,d\}$ to $\{1,2,\cdots,d\}$ with the property that $1 \leq f_n(j) \leq j$. Let $\delta_{s,t} = 1\ (\textmd{if}\ s=t)$ or $0\ (\textmd{if}\ s\neq t)$.
Then there is a permutation $\pi_n$ such that
\bea
K_n \ =  P_{\pi_n}\small{\left(
                          \begin{array}{ccccc}
                            k_1^{(n)} & \delta_{1,f_n(2)}k_2^{(n)} & \cdots & \delta_{1,f_n(d-1)}k_{d-1}^{(n)} & \delta_{1,f_n(d)}k_d^{(n)} \\
                            & & & &\\
                            0 & \delta_{2,f_n(2)}k_2^{(n)} & \cdots & \delta_{2,f_n(d-1)}k_{d-1}^{(n)} & \delta_{2,f_n(d)}k_d^{(n)} \\
                            & & & &\\
                            0 & 0 & \cdots & \delta_{3,f_n(d-1)}k_{d-1}^{(n)} & \delta_{3,f_n(d)}k_d^{(n)} \\
                            & & & &\\
                            \vdots & \vdots & \ddots  & \vdots & \vdots \\
                            & & & &\\
                            0 & 0 & \cdots & 0 & \delta_{d,f_n(d)}k_d^{(n)} \\
                          \end{array}
                        \right)}.
\eea
From $\sum_n K_n^\dagger K_n = \id$, we get that
\bea\label{1-4}
\left\{
                          \begin{array}{l}
                          \sum_n |k_j^{(n)}|^2 = 1 \quad (j=1,2,\cdots,d),\\
                          \\
                          \sum_n \overline{k_1^{(n)}}\delta_{1,f_n(j)}k_j^{(n)} = 0 \quad (j=2,\cdots,d),\\
                          \\
                          \sum_n \sum_l \overline{k_i^{(n)}}k_j^{(n)}\delta_{l,f_n(i)}\delta_{l,f_n(j)} = 0 \\
                          \end{array}
                        \right.                         \qquad \quad  (2\leq i<j\leq d \ \ \textrm{and} \ \ l=1,2,\cdots,i).
\eea

We can see from the definition of maximum fidelity $\rF$ that, for any $n$, $\rF$ increases when all $k^{(n)}_i$ in the matrix $K_n$ are placed in different rows. Therefore, we can prove the following without any loss, assuming that all $k^{(n)}_i$ are arranged in different rows each other.

Returning to our main purpose, we prove the monotonicity for the case when $\rho$ is an arbitrary pure state.
It is straightforward to verify that if $\rho$ is a pure state, then $\rho_n = K_n\rho K_n^\dagger/p_n$ is also a pure state, where $p_n = \tr{K_n\rho K_n^\dagger}$, and from Ineq.(\ref{1-3}) and the above assumptions, we obtain $\rF_{\cC}(\rho_n) \leq \frac{1}{d}\sum_{i,j}|k^{(n)}_{ij}\rho_{ij}|/p_n$. For any $i,j$, we also have
\begin{eqnarray}\label{1-5}
 \sum_n |k^{(n)}_{ij}| \leq \frac{\sum_n \big\{|k^{(n)}_i|^2+|k^{(n)}_j|^2\big\}}{2} = 1\ \
\end{eqnarray}
where $k^{(n)}_{ij}= k^{(n)}_i\overline{k^{(n)}_j}$, and it implies that
\begin{eqnarray}\label{1-6}
\sum_np_n\rF_{\cC}(\rho_n) &\leq& \frac{1}{d}\sum_n \sum_{i,j}|k^{(n)}_{ij}\rho_{ij}|
 =  \frac{1}{d}\sum_{i,j}\big(\sum_n |k^{(n)}_{ij}|\big)|\rho_{ij}|
 \leq \frac{1}{d}\sum_{i,j}|\rho_{ij}| = \rF_{\cC}(\rho).
\end{eqnarray}
Therefore, we have
\begin{eqnarray}\label{1-7}
\sum_np_n\rD^C_g(\rho_n) = 1- \sum_np_n\rF_{\cC}(\rho_n) \geq  1- \rF_{\cC}(\rho) = \rD^C_g(\rho).
\end{eqnarray}

\subsection{Proof of Theorem~\ref{thm2}}
As in the case of coherence, this requires only a proof for (D2b) to confirm that $\rD^E_g$ is a measure of deficiency.
By the arbitrary maximally entangled states $\sigma_{\phi} =\out{\phi}{\phi}$ are in the form of $\ket{\phi} = \frac{1}{\sqrt{d}}\sum_{i=0}^{d-1}\ket{\phi_i}_A\ket{\phi_i}_B$ where $\min\{\dim\{\cH_A\}, \dim\{\cH_B\}\}=d$, and $\{\ket{\phi_i}_A\}_i$, $\{\ket{\phi_i}_B\}_i$ are orthogonal states of the systems $A$ and $B$, respectively, they are derived as follows:
\be
\rF_{\cE}(\rho) = \max_{\sigma_{\phi}\in\cE^{\max}}\rF(\sigma,\rho)= \frac{1}{d}\max_{\sigma_{\phi}\in\cE^{\max}}\sum_{i,j=0}^{d-1}\rho^{(\phi)}_{iijj}
\ee
where $\rho^{(\phi)}_{efgh} = \iinner{\phi_{e}|_A\langle\phi_{f}|_B\rho}{\phi_{g}\rangle_A|\phi_h}_B$. Therefore,
\be
\rD^E_g(\rho) = 1-\max_{\sigma_{\phi}\in\cE^{\max}}\frac{1}{d}\sum_{i,j=0}^{d-1}\rho^{(\phi)}_{iijj}.
\ee
If $\rho$ is pure, \emph{i.e.}, $\rho=\out{\psi}{\psi}$ where $\ket{\psi} = \sum_iq_i\ket{\phi_i}_A\ket{\phi_i}_B$ with a positive real $q_i$ for every $i$, we have
\be\label{2-1}
\rD^E_g(\rho) = 1-\frac{1}{d}\sum_{i,j=0}^{d-1}q_iq_j.
\ee

To prove monotonicity for pure states $\rho = \out{\psi}{\psi}$ with $\ket{\psi} = \sum_iq_i\ket{\phi_i}_A\ket{\phi_i}_B$, we first demonstrate monotonicity under local unitary operations. Let $\Phi_U$ is a local unitary operation acted on by unitary operators $U_A$ and $U_B$, \emph{e.i.}, $\Phi_U(\rho) = U_A\ot U_B\rho U_A^\dagger\ot U_B^\dagger$.
Then
\be\label{2-2}
\rD^E_g(\rho)=\rD^E_g\big[\Phi_U(\rho)\big]
\ee
is induced through
\be\label{2-3}
\rF_{\cE}\big[\Phi_U(\rho)\big] = \frac{1}{d}\sum_{i,j}\iinner{\phi'_{i}|_A\langle\phi'_{i}|_B\Phi_U(\rho)}{\phi'_{j}\rangle_A|\phi'_j}_B
\ee
 from the definition of $\rF_{\cE}$, where  $\ket{\phi'_i}_A = U_A\ket{\phi_i}_A$ and $\ket{\phi'_i}_B =U_B\ket{\phi_i}_B$ for any $i$.

Next, for any local operation $\Phi_{A}\ot \Phi_{B}$, acting as
$$\Phi_{A}\ot \Phi_{B}(\rho) = \sum_{n,m} p_{n,m}\rho_{n,m}$$
where $p_{n,m} = \tr{K^{(A)}_{n}\ot K^{(B)}_{m}\rho (K^{(A)}_{n})^\dagger\ot (K^{(B)}_{m})^\dagger}$ and $\rho_{n,m} = \big\{K^{(A)}_{n}\ot K^{(B)}_{m}\rho (K^{(A)}_{n})^\dagger\ot (K^{(B)}_{m})^\dagger\big\}/p_{n,m}$, we prove that $\sum_{n,m} p_{n,m}\rD^E_g(\rho_{n,m})\geq \rD^E_g(\rho)$.

To do this, we first consider local operations on the single system $A$, such that
\be\label{2-4}
\Phi_{A}\ot \id_B(\rho) = \sum_{n} p_{n}\rho_{n}
\ee
where $p_{n} = \trace\big\{(K^{(A)}_{n}\ot I_{B})\rho (K^{(A)}_{n}\ot I_{B})^\dagger\big\}$ and $\rho_{n} = \big\{(K^{(A)}_{n}\ot I_{B})\rho (K^{(A)}_{n}\ot I_{B})^\dagger\big\}/p_{n}$, and prove that $\sum_{n} p_{n}\rF_{\cE}(\rho_{n})\leq \rF_{\cE}(\rho)$.
For a pure state $\rho=\out{\psi}{\psi}$ whose Schmidt decomposition is given by $\ket{\psi}=\sum_i q_i\ket{\phi_i}_A\ket{\phi_i}_B$, if $K^{(A)}_{n}\ot I_{B}\ket{\phi_i}_A\ket{\phi_i}_B = q_{i,n}\ket{\phi_{i,n}}_A\ket{\phi_{i}}_B$, it is immediately apparent that all $\rho_n$ are also pure states. While $\{\ket{\phi_{i,n}}_A\}_i$ are not necessarily orthonormal, we can assume the following Schmidt decomposition for each $\rho_n=\out{\psi_n}{\psi_n}$:
\be\label{2-5}
\ket{\psi_n}=\sum_iq'_{i,n}\ket{\phi'_{i,n}}_A\ket{\phi'_{i,n}}_B.
\ee
Then, for each $n$, the reduced state $\ptr{B}{\rho_{n}}=\rho_{n,A}$ of particle $A$ can be expressed in the following two ways:
\begin{eqnarray}\label{2-6}
\rho_{n,A}=\sum_{i}(q'_{i,n})^2\ket{\phi'_{i,n}}_A\bra{\phi'_{i,n}}=\sum_{i}\frac{q_{i}^2|q_{i,n}|^2}{p_n}\ket{\phi_{i,n}}_A\bra{\phi_{i,n}}.
\end{eqnarray}

Here, we recall that the set $\{\ket{\phi'_{i,n}}_A\}_i$ is orthonormal. This allows us to prove the majorization relation $\{(q'_{i,n})^2\}_i\succ\{\frac{q_{i}^2|q_{i,n}|^2}{p_n}\}_i$. Then, assuming that it does not hold, there exists some $m (<d)$ that satisfies $p_n\sum_{i=1}^m(q'_{i,n})^2<\sum_{i=1}^m q_{i}^2|q_{i,n}|^2$. Let $\{\ket{\varphi_i}\}_{i=1}^l\ (l\leq m)$ be a basis for the subspace of $\cH_A$, the closed linear span of $\{\ket{\phi_{i,n}}_A\}_{i=1}^m$. Then, we can confirm that the following inequality holds:
\begin{eqnarray}
p_n\sum_j^l\bra{\varphi_j}\rho_{n,A}\ket{\varphi_j}&=&\sum_{j}^l\sum_{i}^dq_{i}^2|q_{i,n}|^2|\iinner{\varphi_j}{\phi_{i,n}}_A|^2\nonumber\\
&=&\sum_{i}^mq_{i}^2|q_{i,n}|^2+\sum_{j}^l\sum_{m+1}^dq_{i}^2|q_{i,n}|^2|\iinner{\varphi_j}{\phi_{i,n}}_A|^2\nonumber\\
&\geq& \sum_{i}^mq_{i}^2|q_{i,n}|^2>p_n\sum_{i=1}^m(q'_{i,n})^2.
\end{eqnarray}
This contradicts the fact that, since $l<m$, it must be true that $\sum_j^l\bra{\varphi_j}\rho_{n,A}\ket{\varphi_j} \leq \sum_{i=1}^m(q'_{i,n})^2$ (see \cite{Watrous}). Therefore, $\{(q'_{i,n})^2\}_i\succ\{\frac{q_{i}^2|q_{i,n}|^2}{p_n}\}_i$ holds.
Then, from the Schur-concavity \cite{Bhatia1997} of symmetric concave function $f(x)=\sum_{i,j}\sqrt{x_ix_j}$ for $x=\{x_i\}_i$, this leads to the following inequality:
\be\label{2-7}
\sum_{i,j}\big(q'_{i,n}q'_{j,n}\big) \leq\frac{1}{p_n}\sum_{i,j}\big(q_iq_j|q_{i,n}q_{j,n}|\big).
\ee
The above equality is obtained when the set of vectors $\{\ket{\phi_{i,n}}_A\}_i$ is orthonormal set.
Furthermore, from $\sum_n (K^{(A)}_{n})^\dagger K^{(A)}_{n}=I_A$, it can be inferred that $\sum_n|q_{i,n}|^2 = 1$ for any $i$. Consequently, for every $i, j$, we obtain the following inequality via the Cauchy-Schwarz inequality:
\begin{eqnarray}\label{2-8}
\sum_{n}|q_{i,n}q_{j,n}| \leq \sum_{n}\frac{|q_{i,n}|^2+|q_{j,n}|^2}{2} = 1.
\end{eqnarray}
Therefore, we obtain the following results from Ineqs. (\ref{2-7}) and (\ref{2-8}) :
\begin{eqnarray}\label{2-9}
\sum_{n} p_{n}\rF_{\cE}(\rho_{n}) &=& \frac{1}{d}\sum_{n}p_n\sum_{i,j}\big(q'_{i,n}q'_{j,n}\big)\leq \frac{1}{d}\sum_{n}\sum_{i,j}\big(q_iq_j|q_{i,n}q_{j,n}|\big)\nonumber\\
 &\leq& \frac{1}{d}\sum_{i,j}\big(q_iq_j\sum_{n}\frac{|q_{i,n}|^2+|q_{j,n}|^2}{2}\big)
 = \frac{1}{d}\sum_{i,j}q_iq_j = \rF_{\cE}(\rho).
\end{eqnarray}

Similarly, we establish an inequality $\sum_{m} \frac{p_{n,m}}{p_{n}}\rF_{\cE}(\rho_{n,m})\leq \rF_{\cE}(\rho_{n})$ for each $n$.
In the end, it implies that $\sum_{n,m} p_{n,m}\rD^E_g(\rho_{n,m})\geq \rD^E_g(\rho)$.

\section{Proof of Theorem 3}
\label{Appendix B}

Specifically, for the case $\dim(\mathcal{H})=2$, let $\{\theta_{ij}\}_{i,j}$ be the set of angles for which $\rho_{ij}=e^{i\theta_{ij}}|\rho_{ij}|$ holds for $i,j \in \{0,1\}$.
Since the system is two-dimensional, the only off-diagonal elements are $\rho_{01}$ and $\rho_{10}$.
We can always find a solution set $\{\theta'_0, \theta'_1\}$ for the basis phases satisfying the equations
\begin{eqnarray}
\theta'_1 - \theta'_0 = -\theta_{01}, \quad \theta'_0 - \theta'_1 = -\theta_{10}.
\end{eqnarray}
Note that in two dimensions, $\theta_{10} = -\theta_{01}$ holds automatically, ensuring the consistency of the equations.

Then, for the maximally coherent state $\ket{\psi}=\sum_{i=0}^1 e^{i\theta'_i}\ket{i}$ constructed from this solution, we have
\begin{eqnarray}\label{3-1}
\rF_{\cC}(\rho) \geq \rF(\out{\psi}{\psi},\rho)
&=& \frac{1}{2}\sum_{i,j=0}^1 e^{i(\theta'_j - \theta'_i)}\rho_{ij} \nonumber \\
&=& \frac{1}{2}\sum_{i,j=0}^1 e^{i(\theta'_j - \theta'_i)}e^{i\theta_{ij}}|\rho_{ij}| \nonumber \\
&=& \frac{1}{2}\sum_{i,j=0}^1 |\rho_{ij}|.
\end{eqnarray}
It follows that
\be\label{3-2}
\rD^C_g(\rho) = 1-\frac{\sum_{i,j=0}^1|\rho_{ij}|}{2}
\ee
for any $\rho\in\cD(\cH)$.
Therefore, we can see from Ineq. (\ref{1-6}) that $\rD^C_g$ is a measure of resource deficiency that satisfies the universal monotonicity (U-D2b) for all quantum states $\rho\in\cD(\cH)$ with $\dim(\mathcal{H})=2$.

\section{Maximum probability success in the subchannel discrimination}
\label{Appendix C}

In the subchannel discrimination,  $P_{succ}\big(\{\Psi_i\}, \{M_i\}, \rho\big)=1$ if and only if the measurement $\{M_i\}$ satisfies $\Pi_{\Psi_i(\rho)}\leq M_i$ for each $i$, where $\Pi_\sigma$ is the projection operator over all eigen-states with nonzero eigenvalues for $\sigma$.

\begin{fact}\label{fact1}
If $\sigma$ is any pure state, then  there are countless strategy $(\{\Psi_i\}, \{M_i\})$ that satisfy $P_{succ}\big(\{\Psi_i\}, \{M_i\}, \sigma\big)=1$.
\end{fact}
This is a clear fact and we can give a useful example here. Let $\{\ket{\varphi_i}\}$ is a basis of $\cH$.
Suppose that in a strategy $(\{\Psi_i\}, \{M_i\})$ operations $\{\Psi_i\}$ are defined through a series of unitary operators $U_i$ that perform $\Psi_i(\sigma) = p_iU_i\sigma U_i^\dagger = p_i\out{\varphi_i}{\varphi_i}$ with a probability distribution $(p_i)$. At this time, if measurement $M_i = \out{\varphi_i}{\varphi_i}$ is performed on the quantum states converted through $\{\Psi_i\}$, we have a success probability of $1$ regardless of the probability distribution $(p_i)_i$ in which the operations $\{\Psi_i\}$ are performed.

Then, from the definition of $\Omega_\sigma$, Eq. (9) in main text can be rewritten as
\be\label{31}
\max_{\sigma\in \cS^{\max}}\min_{\Omega_\sigma}P_{succ}(\{\Psi_i\}, \{M_i\}, \rho).
\ee

\section{Proof of Theorem 5}
\label{Appendix D}

We first consider $\min_{\Omega_\sigma}P_{succ}(\{\Psi_i\}, \{M_i\},\rho)$ for the quantum state $\rho$ and for any maximum resource state $\sigma$. Here, all maximum resource states $\sigma$ are pure states and can be written as $\sigma = \out{\phi_\sigma}{\phi_\sigma}$. We already know that the measurement $\{M_i\}$ that satisfies $\Pi_{\Psi_i(\out{\phi_\sigma}{\phi_\sigma})}\leq M_i$ for any strategy $(\{\Psi_i\}, \{M_i\})$.
It implies that, for any pure state $\out{\psi}{\psi}$, the following inequality is established
\be\label{32}
\tr{M_i\Psi_i(\out{\psi}{\psi})} \geq \tr{\Psi_i(\out{\psi}{\psi})}|\iinner{\phi_\sigma}{\psi}|^2
\ee
where $\ket{\psi} = \iinner{\phi_\sigma}{\psi}\ket{\phi_\sigma} + \delta \ket{\phi^\perp_\sigma}$ with $|\iinner{\phi_\sigma}{\psi}|^2+ |\delta|^2 = 1.$
Therefore, when the spectral decomposition of $\rho$ is $\rho = \sum_jq_j\out{\psi_i}{\psi_i}$, we have that
\begin{eqnarray}\label{33}
P_{succ}(\{\Psi_i\}, \{M_i\}, \rho) &=& \sum_{i,j}q_j \tr{M_i\Psi_i(\out{\psi_j}{\psi_j})}\nonumber\\
 &\geq& \sum_{i,j}q_j\tr{\Psi_i(\out{\psi_j}{\psi_j})}|\iinner{\phi_\sigma}{\psi_j}|^2\nonumber\\
 &=& \sum_{j}q_j|\iinner{\phi_\sigma}{\psi_j}|^2 = \rF(\sigma, \rho),
\end{eqnarray}
Since this inequality is established for any strategy $(\{\Psi_i\}, \{M_i\})$ , we obtain that
\be\label{34}
\min_{\Omega_\sigma}P_{succ}(\{\Psi_i\}, \{M_i\}, \rho)\geq \rF(\sigma, \rho).
\ee

Conversely, we can design a strategy for all maximum resource state $\sigma$:
Let $\{\ket{\varphi_i}\}$ is a basis of $\cH$.
Suppose that in a strategy $(\{\Psi'_i\}, \{M'_i\})$, for each $i$, operation $\Psi_i$ implemented is defined through the unitary operator $U_i$ that perform $\Psi'_i(\sigma) = p_iU_i\sigma U_i^\dagger = p_i\out{\varphi_i}{\varphi_i}$ with a probability distribution $(p_i)$, and the measurement $\{M'_i\} = \out{\varphi_i}{\varphi_i}$ is performed on the quantum states converted through $\{\Psi'_i\}$. Then, we have
\begin{eqnarray}\label{35}
P_{succ}(\{\Psi'_i\}, \{M'_i\}, \rho) &=& \sum_{i,j}q_j \tr{\out{\varphi_i}{\varphi_i}\Psi'_i(\out{\psi_j}{\psi_j})}\nonumber\\
&=& \sum_{i,j}q_j \tr{U_i^\dagger\out{\varphi_i}{\varphi_i}U_i U_i^\dagger\Psi'_i(\out{\psi_j}{\psi_j})U_i}\nonumber\\
&=& \sum_{i,j}p_iq_j|\iinner{\phi_\sigma}{\psi_j}|^2 = \rF(\sigma, \rho).
\end{eqnarray}
This means
\be\label{36}
\min_{\Omega_\sigma}P_{succ}(\{\Psi_i\}, \{M_i\}, \rho)= \rF(\sigma, \rho),
\ee
so, we get that
\begin{eqnarray}\label{37}
\max_{\sigma\in \cS^{\max}}\min_{\Omega_\sigma}P_{succ}(\{\Psi_i\}, \{M_i\}, \rho)= \max_{\sigma\in \cS^{\max}}\rF(\sigma, \rho)=  1-\rD_g(\rho).
\end{eqnarray}

\section{Concrete Scheme and Results for Noise-Constant Estimation via Deficiency Measure}
\label{Appendix E}

Assuming that $\Phi_{\epsilon_{\textmd{H}}}^{\otimes n}$ represents the transformation under the noisy environment of $H^{\otimes n}$, we can compute the fidelity between the initial state $\out{0}{0}^{\otimes n}$ and the state $\rho_{2}^{\textmd{noise}} = \Phi_{\epsilon_{\textmd{H}}}^{\otimes n}[\Phi_{\epsilon_{\textmd{H}}}^{\otimes n}(\out{0}{0}^{\otimes n})]$ after two applications of the channel. For a single qubit:
\begin{equation}
\Phi_{\epsilon_{\textmd{H}}}(\rho)= (1-\epsilon_{\textmd{H}})H\rho H^\dagger + \epsilon_{\textmd{H}}I/2
\end{equation}
which is equivalent to a depolarizing channel. Thus,
\begin{equation}
\rho_{2}^{\textmd{noise}} =\big[(1-\epsilon_{\textmd{H}})^2\out{0}{0}+(\epsilon_{\textmd{H}}-\frac{\epsilon_{\textmd{H}}^2}{2})I\big]^{\otimes n}.
\end{equation}
When the fidelity between the two states is defined as $F_{n}=F(\out{0}{0}^{\otimes n},\rho_{2}^{\textmd{noise}})$, we obtain
\begin{equation}
F_{n}=(1-\epsilon_{\textmd{H}}+\frac{\epsilon_{\textmd{H}}^2}{2})^n.
\end{equation}
Therefore, from
\begin{equation}
F_{\textmd{max}}(\rho^{\textmd{noise}}) = F(\out{+}{+}^{\otimes n}, \rho^{\textmd{noise}}) = (1 - \epsilon_{\textmd{H}}/2)^n,
\end{equation}
we obtain the relation
\begin{equation}
F_{\textmd{max}}(\rho^{\textmd{noise}}) = \left[ \frac{1 + \sqrt{2(F_n)^{1/n}-1}}{2} \right]^n .
\end{equation}

To account for additional errors introduced during the actual inner product estimation performed under the same noise environment, we incorporate the following correction. We compute the inner product estimation fidelity $F_{\textmd{meas}}$ under the general assumption that $\epsilon_{\textmd{CNOT}}\approx 10\epsilon_{\textmd{H}}$, where the estimation fidelity is approximated by
\begin{equation}
F_{\textmd{meas}} = (1-\epsilon_{\textmd{H}})^2(1-\epsilon_{\textmd{CSWAP}})^n(1-\epsilon_{\textmd{M}}).
\end{equation}
This computes the estimation fidelity for a conventional inner product estimation algorithm composed of $2$ Hadamard gates in the measurement system and $n$ CSWAP gates for each target system of n-qubit states, and a measurement gate. Implementing a CSWAP gate requires $1$ Toffoli gate and $2$ CNOT gates, and since a single Toffoli gate is typically composed of $6$ CNOT gates, we can consider that
$\epsilon_{\textmd{CSWAP}}\approx 80\epsilon_{\textmd{H}}.$ Furthermore, assuming that the given inner product estimation process operates in an error-corrected environment and that the measurement noise satisfies $\epsilon_M \approx 5\epsilon_\textmd{H}$, then for $\epsilon_{\textmd{H}} \ll 1$, we can approximately express the total noise $\epsilon_{\textmd{total}}$ corresponding to the overall inner product estimation and the estimation fidelity in terms of $\epsilon_{\textmd{H}}$ as follows:
$\epsilon_{\textmd{total}}\approx 2\epsilon_{\textmd{H}}+n(80\epsilon_{\textmd{H}})+ 5\epsilon_\textmd{H},$ and $F_{\textmd{meas}}\approx 1- \epsilon_{\textmd{total}} \approx 1- (80n+7)\epsilon_{\textmd{H}}.$

 Denoting by $F_{\text{measured}}$ the empirically obtained average inner product between the two states $\out{0}{0}^{\otimes n}$ and $\rho_{2}^{\textmd{noise}}$, we obtain that
\begin{equation}\label{C-7}
F_{\text{measured}}\approx F_{\textmd{meas}}F_{n}+(1-F_{\textmd{meas}})F_{\textmd{random}},
\end{equation}
with $F_{\textmd{random}}=1/2$ being the expectation value for a random measurement.
That is, neglecting the $\epsilon_\textmd{H}^2$ term, since $F_{\text{measured}} \approx 1 - \frac{82n + 7}{2} \epsilon_\textmd{H}$, we obtain the approximation
\begin{eqnarray}
F_{\textmd{max}}(\rho^{\textmd{noise}}) \approx \big(1- \frac{1-F_{\text{measured}}}{82n + 7}\big)^n
\approx 1- \frac{n(1-F_{\text{measured}})}{82n + 7} = \frac{n F_{\text{measured}} + 81n + 7}{82n + 7}.
\end{eqnarray}

However, it is important to note that these values are derived from estimates based on typical noise ratios between quantum gates and may not be universally applicable in all experimental scenarios due to variations in hardware implementations and environmental conditions.
Moreover, assuming that intrinsic systematic errors--such as bit-flip errors--have been adequately compensated, we present the error margins for noise estimation at $\epsilon_{\textmd{H}}=5\times10^{-4}$, accounting exclusively for statistical uncertainties.

Since $F_{\text{measured}}=2p_0-1$ ($p_0$ is the probability that the outcome is $0$ among all samples) is obtained as the sample mean of $N_s$ repeated trials, its variance under the binomial sampling model can be approximated, for sufficiently small $\epsilon_\textmd{H}$, by substituting the linearized expression for $F_{\text{measured}}$ into the variance formula for a proportion:
\begin{equation}
    \sigma^2_{F_{\text{measured}}} = \frac{1 - F_{\text{measured}}^2}{N_s}\approx \frac{(82n+7)\epsilon_\textmd{H}}{N_s}.
\end{equation}

And from the previously confirmed expression for $F_{\text{max}}$, the deficiency measure $\rD_g(\rho_{\textmd{noise}})$
 is estimated from the measured fidelity $F_{\text{measured}}$ as follows:
\begin{equation}
    \widehat{\rD}_{g}(\rho_{\textmd{noise}}) = \frac{n(1- F_{\text{measured}})}{82n + 7}.
\end{equation}
To determine the uncertainty in $\rD_g(\rho_{\textmd{noise}})$, we apply the error propagation rule:
\begin{equation}
    \sigma_{\rD_g} = \left| \frac{\partial \rD_g}{\partial F_{\text{measured}}} \right| \sigma_{F_{\text{measured}}} = \frac{n}{82n + 7} \sigma_{F_{\text{measured}}}.
\end{equation}
The corresponding margin of error (ME) for a $95\%$ confidence interval (CI) is calculated as:
\begin{equation}
    ME_{\rD_g} = 1.96 \cdot \sigma_{\rD_g} =\frac{1.96n}{82n + 7} \sigma_{F_{\text{measured}}}.
\end{equation}

Alternatively, from the approximate expression
\begin{equation}
    \hat{\epsilon}_\textmd{H} = \frac{2\widehat{\rD}_g(\rho_{\textmd{noise}})}{n},
\end{equation}
and given the standard deviation $\sigma_{\rD_g}$ of $\rD_g(\rho_{\textmd{noise}})$, the uncertainty $\sigma_{\epsilon_\textmd{H}}$ is determined via error propagation as:
\begin{equation}
    \sigma_{\epsilon_\textmd{H}} = \left| \frac{\partial \epsilon_H}{\partial \rD_g} \right| \sigma_{\rD_g}.
\end{equation}
Consequently, the standard deviation and the $95\%$ confidence margin of error for $\epsilon_\textmd{H}$ are:
\begin{equation}
    \sigma_{\epsilon_\textmd{H}} = \frac{2}{n} \sigma_{\rD_g},
\end{equation}
and
\begin{equation}
    ME_{\epsilon_\textmd{H}} = 1.96 \cdot\\
\sigma_{\epsilon_\textmd{H}} = \frac{2\times1.96 }{n} \sigma_{\rD_g}.
\end{equation}
From this, the relative error of $\epsilon_\textmd{H}$ is calculated as follows:
\begin{equation}
    \frac{ME_{\epsilon_\textmd{H}}}{\epsilon_\textmd{H}}\times100(\%)= 196 \sqrt{\frac{4}{(82n+7)N_s\epsilon_\textmd{H}}}(\%).
\end{equation}

\begin{figure}[!t]
\centering
\includegraphics[height=.28 \textheight]{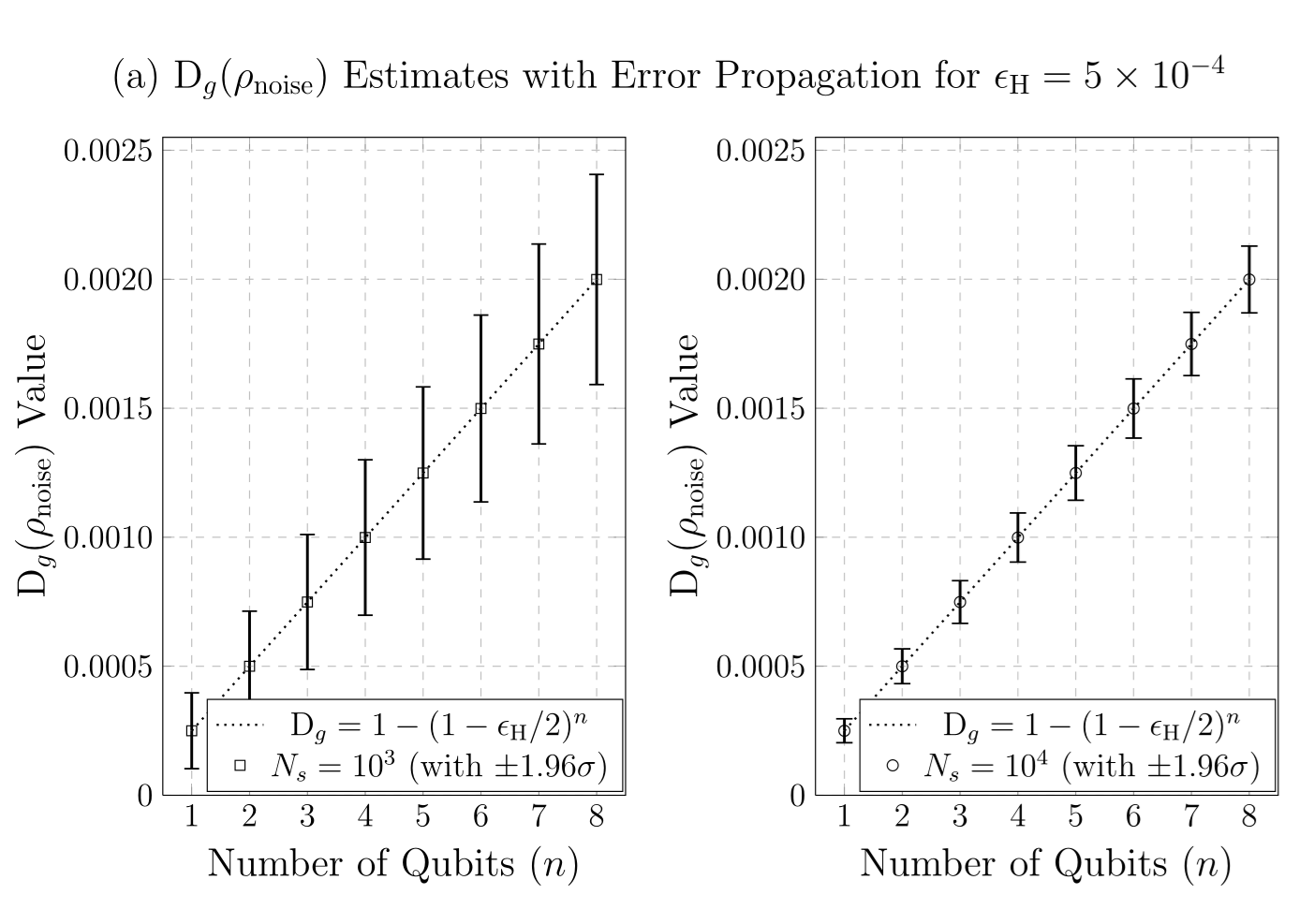}
\includegraphics[height=.265 \textheight]{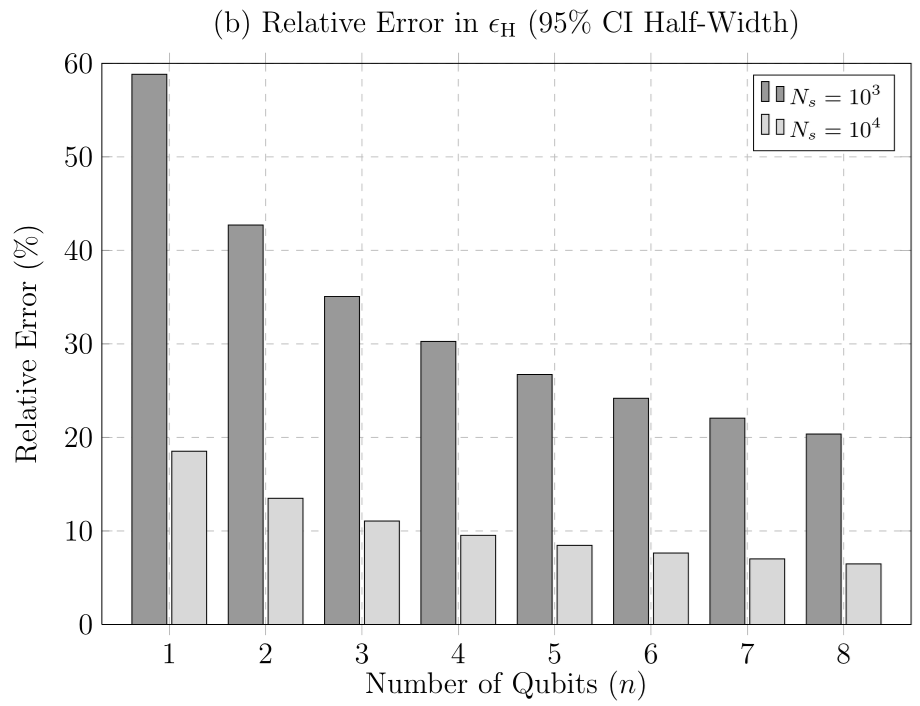}
\caption{\label{fig:3} For each $n$-qubit system with the measurement sample sizes $N_s = 10^3$ or $N_s = 10^4$, we analyze the propagated $95\%$-CI error in the estimation of (a) the deficiency measure $\rD_g(\rho_{\textmd{noise}})$ and (b) the noise constant $\epsilon_{\textmd{H}}$ by applying the error propagation law to the statistical error of the measured average inner product.}
\end{figure}
Based on the same calculation, we obtain Figure \ref{fig:3}, which presents (a) the error margins in estimating the deficiency measure $\rD_g(\rho_{\textmd{noise}})$ and (b) the relative error in estimating the noise constant $\epsilon_\textmd{H}$ for quantum systems with qubit numbers $n = 1,2,\cdots,8$, under measurement sample sizes of $N_s = 10^3$ and $N_s = 10^4$.
Although a sample size of $N_s = 10^3$ yields relative errors exceeding $20\%$, which precludes obtaining meaningful results, the figure demonstrates that for a sample size of $N_s = 10^4$ the relative error in estimating the noise constant drops below $10\%$ (with $95\%$-CI) when the number of qubits $n$ in the quantum system exceeds $4$.

Furthermore, when systematic bias arising from crosstalk is taken into account, the observation model for the inner-product measurement quantifies the crosstalk-error scaling in an $n$-qubit system as follows, based on IBM's Simultaneous Randomized Benchmarking (SRB) methodology \cite{Gambetta} and the error-amplification model due to residual $ZZ$ coupling \cite{McKay}:
\begin{equation}
F'_{\text{meas}} = 1 - (82n + 7) \epsilon_\textmd{H} - \mathcal{E}_\text{ct}^\text{second}+ O(n^3).
\end{equation}
The third term ($\mathcal{E}_\text{ct}^\text{second}$) is an error term added to account for second-order crosstalk as follows
\begin{eqnarray*}
\mathcal{E}_\text{ct}^\text{second}\approx\textmd{(number of executions)}\times\textmd{(concurrency coefficient)}
\times\textmd{(number of adjacent coupling pairs)}\times\textmd{(CNOT weight)}\gamma\epsilon_\textmd{H}.
\end{eqnarray*}
In this estimation, the term ''number of adjacent coupling pairs'' originally reflects the theoretical maximum accumulation of crosstalk that would arise from all possible pairwise interactions. In practice, however, physical-layout constraints reduce the number of coupling pairs between neighboring qubits to approximately $3n$. Moreover, the total number of CNOT executions in the present SWAP-test procedure is $8n$, and the relative strength of second-order crosstalk induced by CNOT-based operations with respect to a single Hadamard noise is quantified by the factor $10\gamma$, where $\gamma$ itself denotes the second-order crosstalk strength, typically on the order of $0.01$ to $0.25$ \cite{Debroy,Dai}.
Due to the complexity of second-order crosstalk, the estimation accuracy of this model degrades rapidly as n increases. As demonstrated in prior studies, gates can be scheduled with staggered execution times, which allows the concurrency coefficient to be reduced to as low as $0.1$ \cite{Yan}. In the inner-product estimation scheme employed here, because CNOT gates are executed in multiples of $8$ proportional to $n$, we set the concurrency coefficient to $1/8$ under ideal scheduling, thereby further suppressing the crosstalk-induced bias in the noise estimation.

Consequently, the second-order crosstalk error term $\mathcal{E}_\text{ct}^\text{second}$ is corrected as the product of the number of executions, the concurrency coefficient, the number of adjacent coupling pairs, and the CNOT weight, yielding $30n^{2}\gamma\epsilon_\text{H}$. This correction transforms an exponentially growing theoretical upper bound into a polynomial-scaling bound that is aligned with realistic experimental conditions.

Also, we neglect third- and higher-order crosstalk terms $(O(n^3))$. The omission is justified by the fact that present-day quantum processors use two-qubit gates as their elementary operations; even the CSWAP gate is implemented as a sequence of two-qubit (CNOT) interactions.
From this, the measured value $F'_{\text{measured}}(n)$ of $F_n$ derived from Eq.~(\ref{C-7}) is as follows:
\begin{equation}
F'_{\text{measured}}(n) \approx 1 - \frac{82n + 7}{2}\epsilon_\textmd{H} - 15n^2\gamma\epsilon_\textmd{H}.
\end{equation}

The accuracy of noise estimation is significantly degraded due to the additional crosstalk strength $\gamma$ that must be taken into account. To estimate $\epsilon_\textmd{H}$, we first employ a method that eliminates $\gamma$; for this purpose, a weighted least-squares (WLS) based linear-regression approach \cite{Carroll} can be utilized. The statistic $\hat{\epsilon}'$ obtained by applying the WLS method to $m$ measurement points is defined as:
\begin{equation}
\hat{\epsilon}'(n_1,\cdots,n_m;d) = \frac{2 (A_{n_1,d}+\cdots+A_{n_m,d})}{B_{n_1,d-1}+\cdots+B_{n_m,d-1}}
\end{equation}
where $A_{n,d}=\frac{1-F'_{\text{measured}}(n)}{(82n+7)^d}$ and $B_{n,d}=\frac{1}{(82n+7)^d}$.
This approach is rooted in the physical fact that ``$\epsilon_\textmd{H}$ is independent of $n$.'' Crucially, the crosstalk-induced bias (the $\gamma$ term) grows with $n$, and this increase in bias also raises the variance of the fidelity measurement, $\sigma^2_{F'_{\text{measured}}}$, as $n$ increases. The WLS estimator assigns weights $\frac{1}{(82n+7)^d}$ that are inversely proportional to the variance of each measurement. Consequently, the difference in the degree of suppression of the bias term in the statistic $\hat{\epsilon}'$ computed for different values of $d$ provides a clue for estimating $\gamma$. In other words, the systematic variation of $\hat{\epsilon}'$ with d can be fitted to effectively back-track $\gamma$. We therefore propose the following estimator for $\gamma$:

\begin{equation}
\hat{\gamma}(n_1,\cdots,n_m;d_1,d_2)=\frac{\hat{\epsilon}'(n_1,\cdots,n_m;d_2)-\hat{\epsilon}'(n_1,\cdots,n_m;d_1)}{30\big(\frac{n^2_1B_{n_1,d_2}
+\cdots+n^2_mB_{n_m,d_2}}{B_{n_1,d_2-1}+\cdots+B_{n_m,d_2-1}}-\frac{n^2_1B_{n_1,d_1}+\cdots+n^2_mB_{n_m,d_1}}{B_{n_1,d_1-1}+\cdots+B_{n_m,d_1-1}}\big)}
\end{equation}

\begin{figure}[!t]
\centering
\includegraphics[height=.265 \textheight]{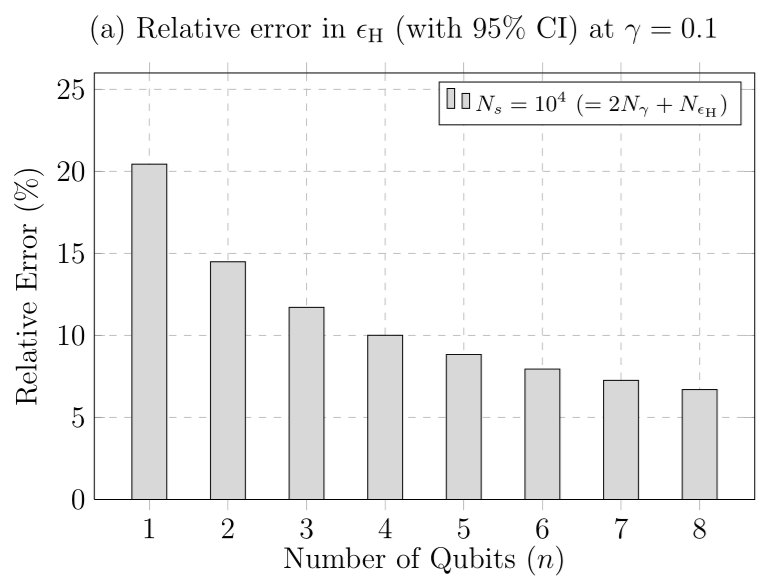}
\includegraphics[height=.28 \textheight]{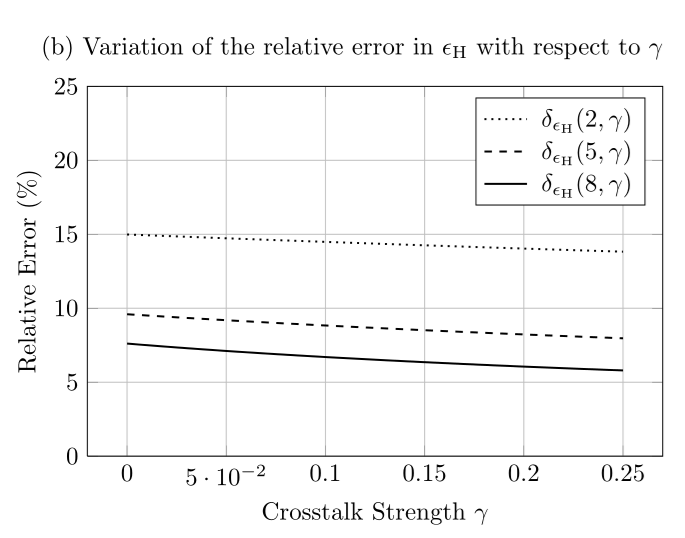}
\caption{\label{fig:4} (a) Relative error in estimating $\epsilon_\text{H}$ versus the number of qubits $n=1,2,\dots,8$ for a fixed crosstalk strength $\gamma=0.1$ and a total sample size of $N_s=10^4$, and (b) its variation as a function of $\gamma \in (0.01, 0.25)$ for selected qubit numbers $n=2,5,8$.}
\end{figure}

For example, with  $\epsilon_\textmd{H} = 5\times10^{-4}$  and a sampling number of  $N_{\gamma} = 10^{3}$  at each point $n_i$, the variance of the estimator $\hat{\gamma}(1,2;2,3)$ is given by
\begin{equation}
\sigma_{\hat{\gamma}}^2 \approx \frac{\epsilon_\textmd{H}}{N_{\gamma}}(0.51 + 0.24\gamma)=2.55\times 10^{-7}+1.2\times 10^{-7}\gamma.
\end{equation}
This shows that, for $\gamma \in (0.01, 0.25)$, the standard deviation  $\sigma_{\hat{\gamma}}$ estimate can be kept within the interval roughly between $3.20 \times 10^{-4}$ and $3.38 \times 10^{-4}$.
Using the estimator of $\gamma$ obtained in this way, we can measure $\rD_g(\rho_{\textmd{noise}})$ via the following estimator:
\begin{equation}
\widehat{\rD}_g(\rho_{\textmd{noise}}) =\frac{n(1-F'_{\text{measured}}(n))}{82n + 7+30n^2\hat{\gamma}}
\end{equation}
Then, the cumulative variance in estimating $\epsilon_{\textmd{H}}$ is given by
\begin{equation}
\sigma_{\widehat{\rD}_g}^2(n) = \frac{n^2\epsilon_{\textmd{H}}}{N_{\epsilon_{\textmd{H}}}(82n+7 + 30n^2\gamma)} + \frac{225n^6\epsilon^2_{\textmd{H}}\sigma_{\hat{\gamma}}^2}{(82n+7 + 30n^2\gamma)^2}.
\end{equation}
For a sample size of $N_{\epsilon_\textmd{H}}=8\times10^{3}$ and $\epsilon_{\textmd{H}}=5\times10^{-4}$, the variance of the propagated $\hat{\epsilon}_\textmd{H}$ is calculated as
\begin{eqnarray}
\sigma_{\hat{\epsilon}_\textmd{H}}^2(n) = \frac{2.5\times 10^{-7}}{82n+7 + 30n^2\gamma} + \frac{2.25\times 10^{-4}n^4\sigma_{\hat{\gamma}}^2}{(82n+7 + 30n^2\gamma)^2}
\approx \frac{2.5\times 10^{-7}}{82n+7 + 30n^2\gamma}.
\end{eqnarray}
The second, approximate expression follows because the second term in the variance formula is negligibly small compared with the first term. And the total sampling number for $\epsilon_\textmd{H}$ estimation becomes $N_s=10^{4}\,(=2N_{\gamma}+N_{\epsilon_\textmd{H}}$, we fix the total number of samples to $10^{4}$ for comparison with the previous results).
Assuming $\gamma=0.1$, the relative error of $\hat{\epsilon}_\textmd{H}$ at the $95\%$ confidence level, defined as
\begin{equation}
\delta_{\epsilon_\textmd{H}}(n;\gamma)(\%)=\frac{1.96\sigma_{\hat{\epsilon}_\textmd{H}}(n)}{\epsilon_\textmd{H}}\times100(\%),
\end{equation}
is plotted against $n=1,2,\dots,8$ in Fig. \ref{fig:4}-(a). Although estimating the crosstalk strength $\gamma$ requires additional experimental resources, inevitably raising the relative error for the same total sample budget $(10^{4})$, we find that the relative error stays below $10\%$ for all $n \ge 5$.

Furthermore, analysis of the relative error formula $\delta_{\epsilon_H}(n;\gamma)$ for $n=2,5,8$ shows that the relative error decreases as the crosstalk strength $\gamma$ increases (Fig. \ref{fig:4}-(b)). This trend is driven by two main factors: First, $\gamma$ appears in the denominator of the leading-order variance term, which reduces the contribution of sampling noise. Second, the effect arising from the uncertainty in estimating $\gamma$ itself is negligible over the range $0.01 \leq \gamma \leq 0.25$.

\section{ProofS of target-repair duality \label{Appendix F}}

\subsection{Proof of Theorem~\ref{thm:6}}

We now relate the average repair capacity to the spectral certificate.
Zero-probability outcomes may be omitted without loss of generality. Independent target optimization on every branch gives
\begin{align}
 \sum_i p_i \rD_{\cT}(\rho_i)
 =1-\sum_i\max_{E_i\in\cT}\tr{E_i\Phi_i(\rho)}
 =1-\max_{\bm E\in\cT^n}
 \tr{\rho\sum_i\Phi_i^\dagger(E_i)},
 \label{eq:F1}
\end{align}
where $\rD_{\cT}(\rho)=1-\max_{E\in\cT}\tr{E\rho}=1-\max_{\omega\in\cC}\tr{\omega\rho}$ is the deficiency function for $\cT$, and $\bm E=(E_1,\dots,E_n)$ denotes the tuple of target effects.

Define the pulled-back operator $X_{\bm E}=\sum_i\Phi_i^\dagger(E_i)$.
By definition, the average repair capacity is
\begin{align}
\Delta \rD_{\cT}(\{\Phi_i\})
 &=\max_{\bm E}\max_{\rho\in\cD(\cH)}
 \min_{\omega\in\cC}\tr{\rho(X_{\bm E}-\omega)}.
 \label{eq:minimax-start}
\end{align}
Since the state space $\cD(\cH)$ and the target hull $\cC$ are both compact and convex, and the objective $\tr{\rho(X_{\bm E}-\omega)}$ is bilinear (hence concave in $\rho$ and convex in $\omega$), Sion's minimax theorem \cite{Sion1958} applies. Exchanging the order of optimization yields
\begin{align}
\Delta \rD_{\cT}(\{\Phi_i\})
 &=\max_{\bm E}\min_{\omega\in\cC}
 \max_{\rho\in\cD(\cH)}\tr{\rho(X_{\bm E}-\omega)}.
 \label{eq:minimax-exchange}
\end{align}
For any fixed Hermitian matrix $A$, the maximization $\max_{\rho\in\cD(\cH)}\tr{\rho A}$ is achieved by the projector onto the eigenspace corresponding to the largest eigenvalue, giving $\lamax(A)$. Consequently,
\begin{align}
\Delta \rD_{\cT}(\{\Phi_i\})
 &=\max_{\bm E}\min_{\omega\in\cC}
 \lamax(X_{\bm E}-\omega).
 \label{eq:instrumentminimax-app}
\end{align}

Because $\Delta \rD_{\cT}(\{\Phi_i\})\geq0$ by definition, inserting a positive-part projection inside the outer maximum does not alter the value:
\begin{equation}
\Delta \rD_{\cT}(\{\Phi_i\})
 =\max_{\bm E}\Bigl[\min_{\omega\in\cC}
 \lamax(X_{\bm E}-\omega)\Bigr]_+.
 \label{eq:positive-part}
\end{equation}
This coincides exactly with the spectral certificate $\mathfrak{S}_{\cT}(\{\Phi_i\})$, thereby proving \eqref{eq:14}.

Next, we describe the order-theoretic foundation underlying \cref{thm:6} to establish the equivalence between Eqs.~\eqref{eq:15} and \eqref{eq:16}.
Let $\cC\subset\textmd{Herm}(\cH)$ be compact and convex, and define the upward closure
\begin{equation}
 \mathord{\downarrow} \mathcal{C} \;=\; \{X\in\textmd{Herm}(\cH): X\leq \omega \text{ for some } \omega\in\cC\}.
\end{equation}
Then, for any Hermitian operator $X$, the following equivalence holds:
\begin{equation}
 \tr{\rho X}\;\leq\;\max_{\omega\in\cC}\tr{\rho\omega}
 \quad\forall\rho\in\cD(\cH)
 \label{eq:F7}
\end{equation}
if and only if $X\in\mathord{\downarrow} \mathcal{C}$.

One direction is immediate: if $X\leq\omega$ for some $\omega\in\cC$, then for every state $\rho\geq0$,
\begin{equation}
 \tr{\rho X}\;\leq\;\tr{\rho\omega}\;\leq\;\max_{\omega'\in\cC}\tr{\rho\omega'},
\end{equation}
which establishes \eqref{eq:F7}.

For the converse, suppose $X\notin\mathord{\downarrow} \mathcal{C}$.
Consider the set $\cC-\Pos = \{\omega-Y : \omega\in\cC,\, Y\geq0\}$.
We first show that $\cC-\Pos$ is closed.
Let $\omega_n-Y_n\to Z$ in norm. Since $\cC$ is compact, $\{\omega_n\}$ admits a convergent subsequence $\omega_{n_k}\to\omega\in\cC$. Then $Y_{n_k}=\omega_{n_k}-(\omega_{n_k}-Y_{n_k})$ also converges to some $Y$. Because the positive semidefinite cone $\Pos$ is closed, $Y\geq0$, so $Z=\omega-Y\in\cC-\Pos$.

Since $X\notin\cC-\Pos$ and $\cC-\Pos$ is closed and convex, the strong separation theorem provides a nonzero Hermitian operator $H$ such that
\begin{equation}
 \tr{HX}\;>\;\sup_{\omega\in\cC,\,Y\geq0}\tr{H(\omega-Y)}.
 \label{eq:strong-separation}
\end{equation}
The right-hand side is finite only if $H\geq0$. Indeed, if $H$ had a negative eigenvalue, we could choose $Y=t\proj{\psi}$ with $\psi$ in the corresponding eigenspace and $t\to\infty$, causing the right-hand side to diverge to $+\infty$. Normalizing $H$ to a density operator $\rho_H=H/\tr{H}\geq0$ would then yield
\begin{equation}
 \tr{\rho_H X}\;>\;\sup_{\omega\in\cC}\tr{\rho_H\omega},
\end{equation}
contradicting \eqref{eq:F7}. Hence $X\in\mathord{\downarrow} \mathcal{C}$, completing the proof.

Consequently, combined with Eq.~\eqref{eq:F1}, this order-theoretic result establishes the equivalence between the nondecreasing deficiency condition \eqref{eq:15} and the adjoint criterion \eqref{eq:16}. Specifically, \eqref{eq:15} holds for all inputs if and only if $X_{\bm E}\in\mathord{\downarrow} \mathcal{C}$ for every admissible tuple $\bm E$, in which case the indicator $\mathfrak{S}_{\cT}(\{\Phi_i\})$ vanishes identically.

\subsection{Proof of Corollary~\ref{cor:7}}
\label{app:continuity-proof}

\textbf{Finite-target SDP (Eq.~\eqref{eq:19}):}
When the target set is finite, $\cT=\{E_1,\dots,E_K\}$, any state in its convex hull can be written as $\omega=\sum_j q_jE_j$ with $\bm q\in\Delta_K$.
For a fixed effect $E_k$, the inner optimization in the repair capacity is
\begin{equation}
 r_k
 =\min_{\omega\in\conv\cT}\,
 \lambda_{\max}\!\bigl(\Lambda^\dagger(E_k)-\omega\bigr).
\end{equation}
Using the spectral characterization
$\lambda_{\max}(H)\leq s\;\Leftrightarrow\;H\leq sI$,
this is equivalent to the semidefinite program
\begin{equation}
 r_k=\min_{s,\bm q}\ s
 \quad\text{s.t.}\quad
 \Lambda^\dagger(E_k)-\sum_jq_jE_j\leq sI,
 \quad \bm q\in\Delta_K,
\end{equation}
which is exactly \eqref{eq:19}.

\textbf{Diamond-norm continuity (Eq.~\eqref{eq:20}):}
Define the affinity $f_{\cT}(\rho)=\max_{\omega\in\cC}\tr{\omega\rho}$.
Since $f_{\cT}$ is linear in $\rho$ and its range is contained in $[0,1]$,
it is 1-Lipschitz with respect to the trace norm:
\begin{equation}
 \bigl|f_{\cT}(\Lambda\rho)-f_{\cT}(\Gamma\rho)\bigr|
 \leq \frac12\|(\Lambda-\Gamma)(\rho)\|_1
 \qquad\forall\,\rho\in\cD(\cH).
\end{equation}
The repair capacity is
$\Delta \rD_{\cT}(\Lambda)=\max_\rho\bigl[D_{\cT}(\rho)-D_{\cT}(\Lambda\rho)\bigr]$,
where $D_{\cT}(\rho)=1-f_{\cT}(\rho)$.
For any input state $\rho$,
\begin{align}
 \Delta \rD_{\cT}(\Lambda)
 &\geq D_{\cT}(\rho)-D_{\cT}(\Lambda\rho)
 = f_{\cT}(\Lambda\rho)-f_{\cT}(\rho),\\
 \Delta \rD_{\cT}(\Gamma)
 &\geq D_{\cT}(\rho)-D_{\cT}(\Gamma\rho)
 = f_{\cT}(\Gamma\rho)-f_{\cT}(\rho).
\end{align}
Subtracting the two inequalities and applying the triangle inequality yields
\begin{align}
 \bigl|\Delta \rD_{\cT}(\Lambda)-\Delta \rD_{\cT}(\Gamma)\bigr|
 \leq \max_\rho \bigl|f_{\cT}(\Lambda\rho)-f_{\cT}(\Gamma\rho)\bigr|
\leq \frac12\max_\rho\|(\Lambda-\Gamma)(\rho)\|_1
 = \frac12\|\Lambda-\Gamma\|_\diamond,
\end{align}
which is \eqref{eq:20}.

\textbf{Deficiency budget (Eq.~\eqref{eq:21}):}
By definition, $\Delta \rD_{\cT}(\Lambda)\geq D_{\cT}(\rho)-D_{\cT}(\Lambda(\rho))$ for all $\rho$.
Rearranging this inequality directly gives the operational budget constraint
\begin{equation}
 D_{\cT}(\Lambda(\rho))
 \geq D_{\cT}(\rho)-\Delta \rD_{\cT}(\Lambda),
\end{equation}
which is \eqref{eq:21}.

\textbf{Composition bound (Eq.~\eqref{eq:22}):}
For any intermediate state $\sigma=\Lambda_1(\rho)$,
the definition of the repair capacity implies
\begin{align}
 \Delta \rD_{\cT}(\Lambda_1)
 &\geq D_{\cT}(\rho)-D_{\cT}(\sigma),\\
 \Delta \rD_{\cT}(\Lambda_2)
 &\geq D_{\cT}(\sigma)-D_{\cT}(\Lambda_2(\sigma)).
\end{align}
Adding these two inequalities gives
\begin{equation}
 \Delta \rD_{\cT}(\Lambda_1)+\Delta \rD_{\cT}(\Lambda_2)
 \geq D_{\cT}(\rho)-D_{\cT}(\Lambda_2\!\circ\!\Lambda_1(\rho)).
\end{equation}
Since this holds for every $\rho$, maximizing over $\rho$ on the right-hand side yields
\begin{equation}
 \Delta \rD_{\cT}(\Lambda_2\!\circ\!\Lambda_1)
 \leq \Delta \rD_{\cT}(\Lambda_1)+\Delta \rD_{\cT}(\Lambda_2),
\end{equation}
which is \eqref{eq:22}.

\textbf{Diamond bound via nonrepairing channels (Eq.~\eqref{eq:23}):}
Let $\cN_{\cT}=\{\Gamma:\Delta \rD_{\cT}(\Gamma)=0\}$ be the set of nonrepairing channels.
For any $\Gamma\in\cN_{\cT}$, we have $\Delta \rD_{\cT}(\Gamma)=0$.
Substituting this into the diamond-norm continuity established in Eq.~\eqref{eq:20} yields
\begin{equation*}
 |\Delta \rD_{\cT}(\Lambda)-0|
 = |\Delta \rD_{\cT}(\Lambda)-\Delta \rD_{\cT}(\Gamma)|
 \leq \frac12\|\Lambda-\Gamma\|_\diamond.
\end{equation*}
Since this inequality holds for every $\Gamma\in\cN_{\cT}$,
taking the infimum over all such $\Gamma$ gives the desired upper bound
\begin{equation*}
 \Delta \rD_{\cT}(\Lambda)
 \leq \frac12\inf_{\Gamma\in\cN_{\cT}}\|\Lambda-\Gamma\|_\diamond,
\end{equation*}
which completes the proof of \eqref{eq:23}.

\section{Proof of \cref{thm:8}}
\label{Appendix G}

Let $X_\Lambda$ denote the normalized Choi state of $\Lambda$, defined as the pulled-back Bell operator acting on the maximally entangled state $\Phi_d = \out{\Phi_d}{\Phi_d}$:
\begin{equation}
 X_\Lambda = (\Lambda^\dagger \otimes \id)(\Phi_d).
 \label{eq:choi-def}
\end{equation}
This operator can be identified with the normalized Choi state of the complex conjugate channel $\overline\Lambda$. Specifically, letting $\{K_j\}$ be a set of Kraus operators for $\Lambda$, we have
\begin{equation}
 X_\Lambda = J(\overline\Lambda),
 \qquad
 \overline\Lambda(\rho) = \sum_j K_j^* \rho K_j^{\mathsf T},
 \label{eq:conjugateChoi}
\end{equation}
where $J(\cdot)$ denotes the normalized Choi-Jamio{\l}kowski isomorphism.
Complex conjugation preserves the mixed-unitarity property, so that $X_\Lambda$ and every $\omega \in \cM_d$ are valid normalized Choi states (i.e., positive semidefinite with unit trace).

For any unitary $U$, the target set $\cM_d$ is invariant under conjugation by $(I \otimes U)$. Consequently, every target in the outer optimization of the dual expression yields the identical spectral minimum, reducing the problem to a minimization over $\cM_d$.

Recall from Theorem~\ref{thm:6} that the optimal deficiency gap (repair capacity) is given by
\begin{equation}
 \Delta \rD_{\textsf{ME}}(\Lambda)
 = \min_{\omega \in \conv(\cT_{\rm ME}^{(d)})}
 \lamax(X_\Lambda - \omega).
 \label{eq:G3}
\end{equation}
Since $\conv(\cT_{\rm ME}^{(d)}) = \cM_d$ coincides with the set of normalized Choi states of $d$-dimensional mixed-unitary channels, and since $X_\Lambda$ is itself a valid normalized Choi state as established above, \eqref{eq:G3} rewrites as
\begin{equation}
 \Delta \rD_{\textsf{ME}}(\Lambda)
 = \min_{\omega \in \cM_d} \lamax(X_\Lambda - \omega),
 \label{eq:G4}
\end{equation}
which is exactly \eqref{eq:30}.

\medskip
\textbf{Equivalence $\Delta \rD_{\textsf{ME}}(\Lambda)=0 \iff \Lambda$ mixed unitary:}
The forward implication is immediate: if $\Lambda$ is mixed unitary, then $X_\Lambda \in \cM_d$. Choosing $\omega = X_\Lambda$ in \eqref{eq:G4} yields $\lamax(0) = 0$, hence $\Delta \rD_{\textsf{ME}}(\Lambda) = 0$.

Conversely, suppose $\Delta \rD_{\textsf{ME}}(\Lambda) = 0$. By \eqref{eq:G4}, there exists an optimal $\omega^* \in \cM_d$ such that $\lamax(X_\Lambda - \omega^*) = 0$.
Since $X_\Lambda - \omega^*$ is Hermitian, $\lamax(X_\Lambda - \omega^*) = 0$ implies $X_\Lambda - \omega^* \le 0$ in the Loewner order (i.e., $\omega^* - X_\Lambda \ge 0$).
Both $X_\Lambda$ and $\omega^*$ are normalized Choi operators (trace one, positive semidefinite). Therefore,
$\tr{\omega^* - X_\Lambda} = 1 - 1 = 0.$
Because $\omega^* - X_\Lambda$ is positive semidefinite and has zero trace, it must be the zero operator. Hence $X_\Lambda = \omega^* \in \cM_d$, proving that $\Lambda$ is mixed unitary. This confirms \eqref{eq:31}

\medskip
\textbf{Choi- and diamond-norm distance bounds:}
The distance bounds follow from a standard spectral inequality.
For any traceless Hermitian operator $H$ acting on a $D = d^2$-dimensional space, the largest eigenvalue is bounded by the trace norm.
Then $\tr{H_+} = \|H\|_1/2$, where $H_+$ is the positive part of $H$.
Since $H$ is traceless, its positive part has rank at most $D-1$; consequently,
\begin{equation}
 \frac{\|H\|_1}{2(D-1)} \leq \lamax(H) \leq \frac{\|H\|_1}{2}.
 \label{eq:G5}
\end{equation}

Let $\Omega_* \in \mathsf{MU}_d$ be the closest mixed-unitary channel to $\Lambda$ in the respective norm.
Recall that $X_\Lambda = J(\overline\Lambda)$ is the normalized Choi state of the complex conjugate channel, so that $H = X_\Lambda - \omega$ is traceless for any $\omega \in \cM_d$.
Applying the lower bound of \eqref{eq:G5} to $H = X_\Lambda - J(\Omega_*)$ and using the fact that $\lamax(A) \le \frac{1}{2}\|A\|_1$ yield
\begin{align}
 2\Delta \rD_{\textsf{ME}}(\Lambda)
 = 2\min_{\omega \in \cM_d} \lamax(X_\Lambda - \omega)
  \le \|X_\Lambda - J(\Omega_*)\|_1
 = \|J(\Lambda) - J(\Omega_*)\|_1
  = \delta_J(\Lambda), \label{eq:choidist-lower}
\end{align}
which is the lower bound in \eqref{eq:33}.

For the upper bound, let $\omega_* \in \cM_d$ attain the minimum in \eqref{eq:G4}, so that $\lamax(X_\Lambda - \omega_*) = \Delta \rD_{\textsf{ME}}(\Lambda)$.
The spectral gap is bounded by the diameter of $\cM_d$ in trace norm. Since $\cM_d$ is the convex hull of the $d^2$-dimensional unitary Choi orbit, its diameter is at most $2(d^2-1)$.
Substituting the upper bound of \eqref{eq:G5} with $D = d^2$ gives
\begin{align}
 \delta_J(\Lambda)
 &\le \|J(\Lambda) - \omega_*\|_1
  \le 2(d^2-1)\, \lamax(X_\Lambda - \omega_*)
  = 2(d^2-1)\, \Delta \rD_{\textsf{ME}}(\Lambda), \label{eq:choidist-upper}
\end{align}
establishing \eqref{eq:33}.

The diamond-norm bounds \eqref{eq:34} then follow immediately from the standard normalized Choi-diamond norm inequalities \cite{Watrous}:
\begin{equation}
 \|J(\Xi)\|_1 \le \|\Xi\|_\diamond \le d\|J(\Xi)\|_1.
 \label{eq:choidiamond-app}
\end{equation}
Complex conjugation preserves both norms and the mixed-unitary set, mapping feasible system-ancilla inputs bijectively. Applying this to the distance measures yields
\begin{align}
 2\Delta \rD_{\textsf{ME}}(\Lambda)
 &\le \|\Lambda - \Omega_*\|_\diamond
  = \delta_\diamond(\Lambda), \\
 \delta_\diamond(\Lambda)
 &\le d\, \delta_J(\Lambda)
  \le 2d(d^2-1)\, \Delta \rD_{\textsf{ME}}(\Lambda),
\end{align}
which proves \eqref{eq:34}.

\section{Proof of \cref{thm:9}}
\label{Appendix H}

\begin{proof}
Define the qubit spin flip (antiunitary involution)
\begin{equation}
\Theta(H)=\sigma_yH^{\mathsf T}\sigma_y,
\qquad
\widetilde\Lambda=\Theta\circ\Lambda\circ\Theta .
\end{equation}
Although $\Theta$ itself is not completely positive, the double conjugation is a channel: if $\{K_j\}$ are Kraus operators of $\Lambda$, then $\{\sigma_yK_j^*\sigma_y\}$ are Kraus operators of $\widetilde\Lambda$. On Bloch vectors,
$\widetilde\Lambda:\bm r\longmapsto M\bm r-\bm t.$
Thus the channel
$\Lambda_0=\frac{\Lambda+\widetilde\Lambda}{2}$
is a unital qubit channel with affine action $\bm r\mapsto M\bm r$. By the Birkhoff theorem~\cite{MendlWolf2009}, every unital qubit channel is mixed unitary. Since mixed-unitary channels cannot increase the fully entangled fraction, they also cannot decrease the deficiency.

Write a two-qubit state $\rho$ in Bloch form, with reference Bloch vector $\bm b$ and correlation matrix $C$.
The channel $\Lambda$ acts on the first qubit via the affine transformation $\bm r \mapsto M\bm r + \bm t$.
This action has two corresponding manifestations depending on the representation space:
\begin{itemize}
    \item Bloch vector space (local action): The Bloch vector $\bm b$ of the first qubit transforms as
    \[
    \bm b \longmapsto M\bm b + \bm t.
    \]
    \item Correlation matrix space (global action): The correlation matrix $C$ transforms as
    \[
    C \longmapsto MC + \bm t\bm b^{\mathsf T}.
    \]
\end{itemize}
Here, the linear part $MC$ represents the transformation under the unital channel $\Lambda_0$, while the rank-one term $\bm t\bm b^{\mathsf T}$ is the perturbation induced by the translation vector $\bm t$ coupling to the second qubit's reference vector $\bm b$.
These two expressions describe the same physical process---the application of $\Lambda$---in the local and global representations, respectively.

In terms of deficiency, the orthogonal-Procrustes representation becomes
\begin{equation}
\rD_{\textsf{ME}}(\rho)=\frac14\left[3-
\max_{O\in O(3),\,\det O=-1}\tr{C^{\mathsf T}O}\right].
\label{eq:DMEprocrustes-app}
\end{equation}
Its support function $\max_{O}\tr{C^{\mathsf T}O}$ is $1$-Lipschitz in the nuclear norm, i.e., for any two matrices $X$ and $Y$, the support function satisfies
\begin{equation}
\Bigl| \max_{O\in\mathcal{K}}\tr{X^{\mathsf T}O} - \max_{O\in\mathcal{K}}\tr{Y^{\mathsf T}O} \Bigr| \leq \|X - Y\|_*,
\end{equation}
where $\mathcal{K} = \{O \in O(3) \mid \det O = -1\}$.
Therefore, the change in deficiency under the rank-one perturbation $\bm t\bm b^{\mathsf T}$ is bounded by
\begin{equation}
\rD_{\textsf{ME}}\!\left[(\Lambda_0\otimes\id)(\rho)\right] - \rD_{\textsf{ME}}\!\left[(\Lambda\otimes\id)(\rho)\right]
\leq\frac14\|\bm t\bm b^{\mathsf T}\|_*
\leq\frac{\|\bm t\|_2}{4},
\end{equation}
where we used $\|\bm t\bm b^{\mathsf T}\|_*\leq\|\bm t\|_2\|\bm b\|_2\leq\|\bm t\|_2$ for a normalized single-qubit input ($\|\bm b\|_2\leq1$).
Since $\Lambda_0$ cannot decrease the deficiency, we have $\rD_{\textsf{ME}}[(\Lambda_0\otimes\id)(\rho)] \geq \rD_{\textsf{ME}}(\rho)$.
Because this upper bound holds for an arbitrary two-qubit state $\rho$, it yields a universal upper bound on the decrease of deficiency, from which we obtain the upper bound on the repair capacity:
\begin{equation}
\Delta \rD_{\textsf{ME}}(\Lambda) \leq\frac{\|\bm t\|_2}{4}.
\end{equation}

To show optimality, take the product input
\begin{equation}
\rho=\frac I2\otimes\proj{\psi_{\bm t}},
\end{equation}
where $\proj{\psi_{\bm t}}=\frac12(I+\bm t/\|\bm t\|_2\cdot\bm\sigma)$ is chosen so that its Bloch vector aligns with $\bm t$. The initial deficiency of this separable state is $\rD_{\textsf{ME}}(\rho) = 1 - 1/4 = 3/4$.
The channel $\Lambda$ maps the first qubit to a state whose Bloch vector is $\bm t$ (up to normalization), so the output deficiency becomes
\begin{equation}
\rD_{\textsf{ME}}\!\left[(\Lambda\otimes\id)(\rho)\right]
= 1 - \frac{1+\|\bm t\|_2}{4} = \frac{3-\|\bm t\|_2}{4}.
\end{equation}
Hence, the decrease in deficiency is exactly
\begin{equation}
\frac{3}{4} - \frac{3-\|\bm t\|_2}{4}
=\frac{\|\bm t\|_2}{4},
\end{equation}
which matches the upper bound and proves that the maximal increase of $\rD_{\textsf{ME}}$ under $\Lambda$ is $\|\bm t\|_2/4$.
This completes the proof of \cref{thm:9}.
\end{proof}

\section{Proof of \cref{thm:10}}
\label{Appendix I}

Direct Choi evaluation gives
\begin{equation}
 J(W_d)=\frac{I-F}{d(d-1)}=\frac{P_-}{r_-}.
 \label{eq:Choi-Wd}
\end{equation}
Thus, the Choi state of the unital channel $W_d$ is exactly the normalized antisymmetric state $P_-/r_-$.

For any maximally entangled vector $|\Phi_U\rangle=(I\otimes U)|\Phi_d\rangle$, its antisymmetric weight is
\begin{equation}
\langle\Phi_U|P_-|\Phi_U\rangle
 =\frac1d\left\|\frac{U-U^{\mathsf T}}2\right\|_{\rm F}^2.
 \label{eq:antisymweight-app}
\end{equation}
Since $U$ is unitary, $\|U\|_{\rm F}^2=d$ and
\begin{equation}
 \left\|\frac{U-U^{\mathsf T}}2\right\|_{\rm F}^2
 = \frac12\bigl(d-\mathrm{Re}\,\tr{U^2}\bigr),
\end{equation}
where $|\tr{U^2}|\le d$ follows from the fact that the eigenvalues of a unitary matrix lie on the unit circle \cite{Bhatia1997}.
Equality $|\tr{U^2}|=d$ forces $U^2$ to be a scalar unitary, and combined with maximization of the antisymmetric weight this restricts the extremal case to real orthogonal matrices.

For real orthogonal matrices, the canonical block decomposition into $2\times2$ rotation blocks and possible $1\times1$ blocks is standard \cite{HornJohnson2012}. In even dimensions, the dimension $d=2k$ can be fully covered by $k$ copies of $2\times2$ blocks.
Crucially, a $2\times2$ rotation block $R(\pi)=\begin{pmatrix}-1&0\\0&-1\end{pmatrix}$ (a $\pi$-rotation) is proportional to the identity and satisfies $R(\pi)^{\mathsf T}=R(\pi)=-I_2$, making it perfectly skew-symmetric. Consequently, in even dimensions the antisymmetric projector $P_-$ fully captures the action of these blocks, allowing the antisymmetric weight to saturate at $q(U)=1$.
In odd dimensions a purely skew-symmetric orthogonal matrix cannot exist: an odd-dimensional real skew-symmetric matrix has even rank, while an invertible skew-symmetric matrix would require odd rank, a contradiction \cite{HornJohnson2012}.
Hence at least one dimension must remain fixed (a $1\times1$ trivial block), and the maximal antisymmetric weight becomes
\begin{equation}
 q(U)\leq q_d=
 \begin{cases}1,&d\ \mathrm{even},\\(d-1)/d,&d\ \mathrm{odd}.
 \end{cases}
 \label{eq:qd-bound}
\end{equation}
The bound is attained by direct sums of $2\times2$ rotation (skew-unitary) blocks, with one scalar block left over in odd dimension.
Moreover, $U(d)$ is connected, $q(U)$ is continuous, and it attains both $0$ and $q_d$; hence its image covers the full interval $[0,q_d]$.

Applying the twirling operation
\begin{equation}
\mathcal{T}(\rho)=\int_{U(d)} (U\otimes U)\,\rho\,(U\otimes U)^\dagger\,dU
\end{equation}
leaves the Choi state $J(W_d)$ invariant, since $J(W_d)$ is itself of Werner form and therefore $U\otimes U$-invariant for all $U\in U(d)$ \cite{Werner1989, HolevoWernerChannels}.
Moreover, for any maximally entangled state $|\Phi_U\rangle=(I\otimes U)|\Phi_d\rangle$, the twirled state $\mathcal{T}(\out{\Phi_U}{\Phi_U})$ remains in the convex hull spanned by all such maximally entangled states; the twirl thus maps the maximally entangled hull onto itself.
Because the twirl is a convex-linear, trace-preserving operation, it does not change the optimal value of the dual optimization problem (Theorem~\ref{thm:8}).
Furthermore, under the $U\otimes U$-symmetry, the feasible set of the dual problem collapses to the one-dimensional Werner subspace,
\begin{equation}
 \omega_q=q\frac{P_-}{r_-}+(1-q)\frac{P_+}{r_+},
 \qquad0\leq q\leq q_d.
 \label{eq:I6}
\end{equation}
so that the dual certificate is completely characterized by the single Werner coordinate $q$ \cite{TwirlingChannels}.
For the noisy channel $W_{d,p}=(1-p)W_d+p\Delta_d$, the Choi state is
\begin{equation}
 X_{W_{d,p}}=\alpha_p\frac{P_-}{r_-}
 +(1-\alpha_p)\frac{P_+}{r_+},
 \label{eq:I5}
\end{equation}
where $\alpha_p=1-\frac{d+1}{2d}p$. If $\alpha_p\leq q_d$, the point $X_{W_{d,p}}$ lies entirely within the target hull defined by \eqref{eq:I6}, so the repair capacity vanishes.
Otherwise, the optimal certificate is $\omega_{q_d}$ and the repair capacity is $(\alpha_p-q_d)/r_-$.
This yields the closed-form expression \eqref{eq:41}.

To verify this explicitly, consider the specific input maximally entangled state $\rho=P_-/r_-$.
By definition, the corresponding deficiency is
\begin{equation}
 \rD_{\textsf{ME}}(\rho)=1-\frac{q_d}{r_-}
 =\begin{cases}
  1-\dfrac{2}{d(d-1)}, & d\ \text{even},\\[1.2ex]
  1-\dfrac{2}{d^2}, & d\ \text{odd}.
 \end{cases}
 \label{eq:DME-input}
\end{equation}
Owing to the $U\otimes U$-symmetry, the antisymmetric score $\alpha_p/r_-$ makes the output deficiency $1-\alpha_p/r_-$. The difference $(\alpha_p-q_d)/r_-$ thus saturates the dual value, completing the proof. \hfill$\square$

\section{Proof of Lemma~\ref{lem:11}}
\label{Appendix J}

The GAD channel with interaction time $t$ and temperature parameter $z$ acts on the Bell basis. Its translated Bell operator is $X_{t,z}=\frac14 I\otimes I + \frac14\tilde{X}_{t,z}$, where
\begin{equation}
\tilde{X}_{t,z}=tX\otimes X-tY\otimes Y+t^2Z\otimes Z+(1-t^2)zI\otimes Z.
\end{equation}
A unitary on the untouched reference can be absorbed into the input without changing its entanglement deficiency $\rD_{\textsf{ME}}$, so output maximization over Bell targets can be fixed to $\Phi_2$(see Eq. (\ref{24})). The operator $X_{t,z}$ is invariant under $R_z(\theta)\otimes R_z(-\theta)$. Twirling the input by this group preserves the objective and, by concavity and local-unitary invariance of $\rD_{\textsf{ME}}$, preserves feasibility. A phase choice then reduces any feasible input to the $X$-form
\begin{equation}
\rho_X=\begin{pmatrix}
r_{00}&0&0&w\\
0&r_{01}&0&0\\
0&0&r_{10}&0\\
w&0&0&r_{11}
\end{pmatrix},\qquad w\ge0.
\end{equation}
For this state, the entanglement deficiency is
\begin{equation}\label{eq:J3}
\rD_{\textsf{ME}}(\rho_X)=1-\max\left\{\frac{r_{00}+r_{11}+2w}{2},\frac{r_{01}+r_{10}}2\right\}.
\end{equation}
Since $r_{01}+r_{10}\le1$, the second term never exceeds $1/2$. Thus the deficiency constraint $\rD_{\textsf{ME}}(\rho)\ge1/2$ is entirely governed by the first term, reducing to $r_{00}+r_{11}+2w\le1$.

Put $x=2w$, and $s=r_{00}+r_{11}$.
To maximize the output Bell score, we require the largest possible Bloch component $b_z$ of the reference system.
For the $X$-form state, $b_z = (r_{00}+r_{10}) - (r_{01}+r_{11})$.
Using the normalization $r_{00}+r_{01}+r_{10}+r_{11}=1$ and introducing $s=r_{00}+r_{11}$ and $x=2w$, we rewrite this as
\begin{equation}
b_z = 2r_{00} - 2r_{01} + 1 - 2s.
\end{equation}
Positivity ($\rho\ge0$) imposes $r_{00}r_{11} \ge w^2$. Substituting $r_{11}=s-r_{00}$ yields the quadratic inequality
\begin{equation}
r_{00}^2 - s r_{00} + \frac{x^2}{4} \le 0,
\end{equation}
which bounds $r_{00}$ within the interval $\frac{s \pm \sqrt{s^2 - x^2}}{2}$.
Since $b_z$ is linear and increasing in $r_{00}$ but decreasing in $r_{01}$, it is maximized by saturating the bounds: $r_{01}=0$ and $r_{00} = \frac{s + \sqrt{s^2 - x^2}}{2}$.
Substituting these extremal values gives the tight upper bound
\begin{equation}
b_z \le 1 - s + \sqrt{s^2 - x^2},
\end{equation}
with equality holding when $r_{01}=0$, $r_{10}=1-s$, and $r_{00}r_{11}=w^2$.
We examine the dependence of the output Bell score $L$ on $s$.
The tight upper bound for the Bloch component is $b_z^{\max}(s)=1-s+\sqrt{s^2-x^2}$.

We define the auxiliary Bell score $L$ as the maximized expectation of the interaction term:
\begin{equation}
L := \max_{\rho_X:\,\rD_{\textsf{ME}}(\rho_X)\ge1/2}\tr{\tilde{X}_{t,z}\rho_X}.
\end{equation}
Using the Bloch components $\tr{X\otimes X\,\rho_X}=2x$ and $\tr{Z\otimes Z\,\rho_X}=2s-1$, the trace evaluates to
\begin{equation}
\tr{\tilde{X}_{t,z}\rho_X} = 2tx + t^2(2s-1) + (1-t^2)z b_z.
\end{equation}
From the positivity constraint derived earlier, the $z$-component of the Bloch vector is bounded by
\begin{equation}
b_z^{\max} = x + \sqrt{1-2x}.
\end{equation}
To maximize $L$, we substitute this upper bound $b_z^{\max}$ into the trace expression, which yields the objective function $L(s)$:
\begin{equation}
L(s)=2tx+t^2(2s-1)+(1-t^2)z\bigl(1-s+\sqrt{s^2-x^2}\bigr).
\end{equation}
Differentiating with respect to $s$ yields
\begin{equation}
\frac{\partial L}{\partial s}=2t^2-(1-t^2)z+\frac{(1-t^2)z\,s}{\sqrt{s^2-x^2}}.
\end{equation}
For any feasible state, $s\ge x>0$, so $\frac{s}{\sqrt{s^2-x^2}}\ge1$. Since $0\le t<1$ and $z>0$, we have
\begin{equation}
\frac{\partial L}{\partial s}\;\ge\;2t^2-(1-t^2)z+(1-t^2)z\;=\;2t^2\;\ge\;0.
\end{equation}
Thus $L(s)$ is a non-decreasing function of $s$: increasing $s$ never reduces the Bell score.
Meanwhile, the entanglement deficiency satisfies $\rD_{\textsf{ME}}(\rho_X)=1-\frac{s+x}{2}$.
Requiring the input to be feasible, i.e., $\rD_{\textsf{ME}}\ge\frac12$, is therefore equivalent to
\begin{equation}
1-\frac{s+x}{2}\;\ge\;\frac12
\quad\Longleftrightarrow\quad
s+x\;\le\;1.
\end{equation}
Because $L$ grows with $s$ and is bounded above by $s\le1-x$, the maximum of $L$ under this constraint is uniquely achieved when the inequality saturates, i.e., $s=1-x.$
This saturation condition will be used in the next step to eliminate $s$ and express the optimal score solely in terms of $x$ (or equivalently $u$).
Write $u=\sqrt{1-2x}$, so $x=(1-u^2)/2$ and the maximal $b_z^{\max}=x+u$. Then
\begin{equation}
L(u)=t+(t^2-t)u^2+\frac{(1-t^2)z}{2}(1+2u-u^2).
\end{equation}
This is strictly concave in $u$ (coefficient $t^2-t<0$). Setting $\partial_u L=0$ yields the unique maximizer
\begin{equation}
u=\frac{(1+t)z}{2t+(1+t)z}.
\end{equation}
Substituting it into $L(u)$ and simplifying algebraically, we obtain
\begin{equation}
L = t + (1-t^2)z\frac{t+(1+t)z}{2t+(1+t)z}.
\end{equation}

For any output state, the entanglement deficiency is related to the Bell expectation by $\rD_{\textsf{ME}} = 1 - \langle X_{t,z}\rangle$.
Therefore, minimizing the deficiency is equivalent to maximizing the Bell expectation. This gives the direct relation
\begin{equation}
\mathcal{D}(t,z) = 1 - \langle X_{t,z}\rangle_{\max}.
\end{equation}
The full Bell expectation is therefore
\begin{equation}
\langle X_{t,z}\rangle_{\max} = \frac14 + \frac14 L.
\end{equation}
Substituting this into the relation $\mathcal{D}(t,z) = 1 - \langle X_{t,z}\rangle_{\max}$ yields
\begin{equation}
\mathcal{D}(t,z) = 1 - \left(\frac14 + \frac14 L\right) = \frac34 - \frac14 L.
\end{equation}
For notational convenience in stating the main theorem, we introduce the frontier function $\ell(t,z):=L$.
Thus, maximizing the Bell score $L$ is strictly equivalent to minimizing $\mathcal{D}(t,z)$.
Substitution gives the exact frontier relation
\begin{equation}
4(1-\mathcal{D}(t,z))-1=\ell(t,z)=t+(1-t^2)z\frac{t+(1+t)z}{2t+(1+t)z},
\end{equation}
with the continuous value $\ell(0,0)=0$.

The optimal input is the rank-at-most-two state
\begin{equation}
\rho_u=\frac14\begin{pmatrix}
(1+u)^2&0&0&1-u^2\\
0&0&0&0\\
0&0&2(1-u^2)&0\\
1-u^2&0&0&(1-u)^2
\end{pmatrix}.
\end{equation}
Its correlation matrix is $\diag(x,-x,u^2)$, so $\rD_{\textsf{ME}}(\rho_u)=(3-2x-u^2)/4=1/2$ exactly. For $0<t<1$, strict activation at the prescribed interaction time occurs if and only if $\ell(t,z)>1$.

Set $D=2t+(1+t)z$. Algebraic manipulation yields
\begin{equation}
\ell(t,z)-1=\frac{1-t}{D}\bigl[(1+t)^2z^2-(1-t^2)z-2t\bigr].
\end{equation}
For $0<t<1$, we have $1-t>0$ and $D>0$, so the sign of $\ell(t,z)-1$ is determined by the quadratic term in the brackets.
Strict activation ($\mathcal{D}\le 1/2$) requires $\ell(t,z)-1 \ge 0$, which implies
\begin{equation}
(1+t)^2z^2-(1-t^2)z-2t \ge 0.
\end{equation}
Solving the positive quadratic root in $z$ gives the critical thermal boundary
\begin{equation}
z_c(t) = \frac{1-t+\sqrt{1+6t+t^2}}{2(1+t)}.
\end{equation}
Consequently, for the corresponding interaction time $\tau = -\frac{2}{\Gamma} \ln t$, strict activation occurs if and only if $z \ge z_c(t)$, which proves Eq.~(\ref{eq:55}). This completes the proof of Lemma~\ref{lem:11}.

\section{Proof of Theorems~\ref{thrm:12}}
\label{Appendix K}
\medskip
\textbf{Global threshold:}
From Eq.~(\ref{eq:55}), minimizing this boundary over $t\in[0,1]$ gives a minimum at $t=1$, with limit $1/\sqrt{2}$. Thus for $z\le1/\sqrt{2}$, no physical $t$ satisfies $z>z_c(t)$, so $\mathcal{A}(z)=0$. This proves the threshold
\begin{equation}
\mathcal{A}(z)=0\quad\Longleftrightarrow\quad z\le\frac1{\sqrt2}.
\end{equation}

\medskip
\textbf{Optimal interaction variable and global activation :}
 Within the active region $z>1/\sqrt{2}$, the derivative of $\ell$ with respect to $t$ is
\begin{equation}
\partial_t\ell=\frac{2t(1+z)}{[2t+(1+t)z]^2}\bigl[z(2-z)+2(1-z^2)t-z(2+z)t^2\bigr].
\end{equation}
The unique positive stationary point is
\begin{equation}
 t_*(z)=\frac{1-z^2+\sqrt{1+2z^2}}{z(2+z)},\qquad
 \gamma_*(z)=1-t_*^2(z).
 \label{eq:K3}
\end{equation}
Endpoint comparison shows it is the global maximum. The activation gain is $\mathcal{A}(z)=\frac{1}{4}(\ell(t_*(z),z)-1)$. Setting $K(z)=(1+t_*(z))z$ for convenience, from Eq.~\eqref{eq:K3}, we derive the algebraic identity
\begin{equation}
2(K(z)+t_*(z))=K(z)(K(z)+2t_*(z)).
\end{equation}
Substituting $t_*$ into $\ell$ and using this identity simplifies $\ell(t_*(z),z)$ exactly to $\frac{(1-t_*(z))(1+t_*(z))^2z^2+2t_*(z)}{2}$. Hence
\begin{equation}
\mathcal{A}(z)=\frac{(1-t_*(z))\big[(1+t_*(z))^2z^2-2\big]}{8}= \frac{(1-t_*(z))\big(K^2(z)-2\big)}{8}.
\end{equation}
Moreover, since both $K^2(z)=(1+t_*(z))^2z^2$ and $1-t_*(z)$ are strictly monotonically increasing for $1/\sqrt{2}\leq z\leq 1$, and $K^2(1/\sqrt{2})=2$ and $1-t_*(1/\sqrt{2})=0$, it follows that $\mathcal{A}(z)>0$ for $z > 1/\sqrt{2}$.

\medskip
\textbf{Optimal input state:}
At $t=t_*(z)$, it follows from Eq.~\eqref{eq:54} that the relation $u_*(z)=\frac{\sqrt{1+2z^2}+z-1}{2+z}$, which is equivalently written as $K(z)-1$, holds. Substituting this into $\rho_u$ in Eq.~\eqref{eq:53} yields the optimal input for the global problem:
\begin{equation}
 \rho_*(z) = \frac{1}{4}
 \begin{pmatrix}
 (1+u_*(z))^2 & 0 & 0 & 1-u_*^2(z) \\
 0 & 0 & 0 & 0 \\
 0 & 0 & 2(1-u_*^2(z)) & 0 \\
 1-u_*^2(z) & 0 & 0 & (1-u_*(z))^2
 \end{pmatrix},
\end{equation}
This state satisfies $\rD_{\textsf{ME}}(\rho_*(z))=1/2$ (see Eq.~\eqref{eq:J3}) and achieves the maximal activation. This completes the proof of Theorem~\ref{thrm:12}.

\end{widetext}

\end{document}